\documentclass{emulateapj}
\usepackage{apjfonts}
\usepackage{epsfig}

\def\calc{{\cal C}}
\def\calg{{\cal G}}
\def\calu{{\cal U}}

\def\dotsfr{\dot{\cal R}}

\def\kms{km~s$^{-1}$}

\def\rhog{\rho_{\mathrm g}}

\def\vr{V_{\rm R}}
\def\siggas{\Sigma_{\mathrm g}}
\def\siggass{\Sigma_{\mathrm g,1}}
\def\sigsfr{\dot{\Sigma}_{\star}}
\def\sunm{M_{\odot}}
\def\sunmyr{M_{\odot}{\rm yr^{-1}}}
\def\sunmyrkpcc{M_{\odot}{\rm yr^{-1}~kpc^{-2}}}

\def\ergs{\ifmmode {\rm ergs~ s^{-1}} \else {\rm ergs~s^{-1}}\ \fi}
\def\kms{\ifmmode {\rm km~ s^{-1}} \else {\rm km~s^{-1}}\ \fi}

\def\HII{\mbox{H {\sc ii}}}
\def\NII{\mbox{[N {\sc ii}]}}
\def\OI{\mbox{[O {\sc i}]}}
\def\OIII{\mbox{[O {\sc iii}]}}
\def\SII{\mbox{[S {\sc ii}]}}
\def\D{\displaystyle}

\slugcomment{Received 2010 February 24; accepted 2010 October 19}
\journalinfo{The Astrophysical Journal, in press}
\shorttitle{Evolution of gaseous disk viscosity driven by supernova explosion. II.}
\shortauthors{Yan \& Wang}

\begin{document}

\title{Evolution of gaseous disk viscosity driven by supernova explosion. II.
Structure and emissions from star-forming galaxies at high redshift }

\author{Chang-Shuo Yan\altaffilmark{1,2} and Jian-Min Wang\altaffilmark{1,3,4}}
\altaffiltext{1}{Key Laboratory for Particle Astrophysics, Institute of High Energy Physics,
 Chinese Academy of Sciences, 19B Yuquan Road, Beijing 100049, China}

\altaffiltext{2}{Graduate University of Chinese Academy of Sciences, 19A Yuquan
Road, Beijing 100049, China}

\altaffiltext{3}{Theoretical Physics Center for Science Facilities, Chinese Academy of Sciences}

\altaffiltext{4}{Correspondence should be addressed to wangjm@ihep.ac.cn}

\begin{abstract}
High spatial resolution observations show that high redshift galaxies are undergoing intensive
evolution of dynamical structure and morphologies displayed by the H$\alpha$, H$\beta$, $\OIII$
and $\NII$ images. It has been shown that supernova explosion (SNexp) of young massive stars during
star formation epoch, as kinetic feedback to host galaxies, can efficiently excite the turbulent
viscosity. We incorporate the feedback into the dynamical equations through
mass dropout and angular momentum transportation driven by the SNexp-excited turbulent viscosity.
The empirical Kennicutt-Schmidt law is used for star formation rates. We numerically solve the
equations and show that there can be intensive evolution of structure of the gaseous disk. Secular
evolution of the disk shows interesting characteristics that are 1) high viscosity excited by
SNexp can efficiently transport the gas from 10kpc to $\sim 1$kpc forming a stellar disk whereas
a stellar ring forms for the case with low viscosity; 2) starbursts trigger SMBH activity
with a lag $\sim 10^8$yr depending on star formation rates, prompting the joint evolution of
SMBHs and bulges; 3) the velocity dispersion is as high as $\sim 100~\kms$ in the gaseous disk.
These results are likely to vary with the initial mass function (IMF) that the SNexp rates rely on.

Given the initial mass function, we use the GALAXEV
code to compute the spectral evolution of stellar populations based on the dynamical structure.
In order to compare the present models with the observed dynamical structure and images, we use
the incident continuum from the simple stellar synthesis and CLOUDY to calculate emission line
ratios of H$\alpha$, H$\beta$, $\OIII$ and $\NII$, and H$\alpha$ brightness of gas photoionized
by young massive stars formed on the disks. The models can produce the main features of emission
from star forming galaxies. We apply the present model to two  galaxies, BX 389 and BX 482
observed in SINS high$-z$ sample, which are bulge and disk-dominated, respectively. Two successive
rings independently evolving are able to reproduce the main dynamical and emission properties of
the two galaxies, such as, BPT diagram, relation between line ratios and H$\alpha$ brightness.
The observed relation between turbulent velocity and the H$\alpha$ brightness can be explained
by the present model. High viscosity excited by SNexp is able to efficiently transport the gas
into a bulge to maintain high star formation rates, or, to form a stellar ring close enough to
the bulge so that it immigrates into the bulge of its host galaxy. This leads to a fast growing
bulge.  Implications and future work of the present models
have been extensively discussed for galaxy formation in high$-z$ universe.
\end{abstract}


\section{Introduction}
Star forming galaxies at redshift $z\sim 1.5 -3.5$ around the peak of cosmic star formation and quasar
activity have a large variety of dynamical properties and morphologies (F\"orster Schreiber et al. 2006;
Genzel et al. 2006, 2008; Wright et al. 2007, 2009; Bournaud et al. 2008). Four main characteristics different
from the local galaxies are found from the large available high$-z$ samples (Genzel et al. 2008; F\"orster
Schreiber et al. 2009; Lehnert et al. 2009; Lemoine-Busserolle \& Lamareille 2009): 1) higher velocity
dispersion $V_{\rm rot}/\sigma<10$ throughout entire regions of galaxies at high redshift than
$V_{\rm rot}/\sigma\sim 15-20$ in local galaxies (Dib et al. 2006), where $V_{\rm rot}$ is the rotational
velocity and $\sigma$ is the velocity dispersion; 2) high$-z$ galaxies are more irregular and asymmetric
in shape, which is not related with ongoing mergers (Shapiro et al. 2008); 3) they show highly clumpy disks,
typically with a scale of $\lesssim 1$kpc (e.g. Genzel et al. 2008); 4) very high surface brightness of the
recombination lines (e.g. Lehnert et al. 2009). There is no doubt that high$-z$ galaxies are undergoing
most violent formation and evolution.

The "heating" mechanism of high velocity dispersion of clumps remains open, which is thought
potentially to be responsible for the
transportation of angular momentum of gas. Turbulence driven by self-gravity instability is not favored
by evidence that such flows are not able to produce the dispersion and emission via mergers (Lehnert et
al. 2009). However, there is increasing evidence for the turbulence excited by supernova explosion (SNexp)
of young massive stars during star formation from both numerical simulations or theoretical arguments
(Wada \& Norman 2001; Mac Law \& Klessen 2004; Wang et al. 2009b) and observations (Dib et al. 2006;
Lehnert et al. 2009). The strong correlation between the Eddington ratios and specific star formation
rates in type 2 AGNs implies the roles of SNexp in exciting turbulence, which triggers an accretion flow
to central SMBHs (Chen et al. 2009; see also Watabe et al. 2008). In particularly, the
strong correlation between the velocity dispersion and star formation rate density in both local
(Dib et al. 2006) and high$-z$ galaxies (Lehnert et al. 2009) indicates the role of the star formation
in triggering turbulence. We have to note
also that turbulence can enhance star formation in term of the compressed density by shocks (Silk 2005).
Anyway, SNexp inputs energy into interstellar medium causing turbulence somehow.

It has been realized that the star formation itself plays a key role in the formation of galactic
structures. The well-known exponential disks of most S and S0 galaxies (e.g. Freeman 1970; Wevers
et al. 1986) can be explained by that the gaseous disk is evolving like a viscose accretion disk,
and the exponential structure reaches when the infalling timescale equals to the star formation
(Lin \& Pringle 1987; Lu \& Cheng 1991), or by self-regulations of star forming galaxies (Dopita
\& Ryder 1994; Silk 1997). Tracing back
the galaxies to high redshift, they must have been undergoing intensive star formation and similar
to those galaxies at high redshift which have exponential disks today. Wang et al. (2009b; hereafter
Paper I) set up equations describing the self-regulating gaseous disk driven by SNexp feedback
for high redshift galaxies, and, fortunately, find that such a set of complex equations has analytical
solutions if the star formation law is a linear relation with gas density. Paper I shows the
important roles of SNexp in the secular evolution to supply gas to $\sim 1$kpc regions and find that
the prominent feature of star forming galaxies in high$-z$ universe is a natural consequence of
turbulent viscosity driven by SNexp. Very recently, Kumar \& Johnson (2010) extend the study
of the roles of SNexp in the transportation of fueling gas at $\sim 1$ kpc scale and confirm that
SNexp is an efficient way to transport angular momentum outward. The advantage
of this self-regulation is that no external torque is needed to transport gas into inner region.
It is obvious that the dynamical structure and morphology is highly time-dependent, which should
be given by a self-consistent way.

On the other hand, coevolution of SMBHs and galaxies has been generally formulated by the correlations
between SMBH mass and bulge luminosity (Magorrian et al. 1998); or dispersion velocity (Ferrarese \&
Merritt 2000; Gebhardt et al. 2000). Great attempts of phenomenological models have been made (e.g. Granato et
al. 2001; Colpi et al. 2006), however, dynamics of the coevolution is poorly understood so far. On the
side of theoretical expectation, gaseous clumps in galaxies not only undergo star formation, but also
will be partially delivered into the central regions of galaxies by SNexp (Paper I). As a
natural consequence of evolution of clumps, bulges are growing and SMBHs do subsequently.  The time lag
of SMBH activities relative to the starbursts is determined by the star formation itself. In principle,
solutions of the time-dependent equations describing the dynamical structures of gaseous disk can present
the whole story of the coevolution.

In this paper, we extend studies of Paper I in detail for the intensive evolution of dynamical
structure and morphologies of high$-z$ galaxies. In \S2, we use the Kennicutt-Schimidt law to
calculate the star formation rates, and incorporate the rates into the dynamics of the gaseous disk
through the SNexp excited-turbulence viscosity. Growth of bulges and SMBHs are discussed as well as
the time lag of triggering SMBH activities with respect to starbursts. Photoionization
model is presented in \S3 for different geometries and conditions in the evolving gaseous disk.
Detailed applications to two high$-z$ galaxies are given in \S4. We extensively discuss implications
and future improvement of the present model in \S5. Conclusions are drawn in the last section.

\section{Dynamics and evolution of the gaseous disks}
\subsection{Dynamical equations}
Spatial resolution of Integral Field Spectroscopy is $\sim 0.1$kpc at $z\sim 2$, which is the maximum
size of clumps resolved observationally (Genzel et al. 2008; Jones et al. 2010). This size is still
smaller than the
characterized length of galaxies ($\sim 10$ kpc), but the number density of the clumps are dense enough
to approximate the system of clumps as continuum fluid. It is valid to use the equations of continuum
medium, but we treat the gas as clumps in the photoionization model. In Paper I, we derive a series of
equations to describe the secular evolution of a gaseous disk by including feedback of star formation.
We assume that turbulence excited by SNexp of massive stars is driving the angular
momentum outward, and mass dropout is happening due to ongoing star formation. Actually the role
of supernova-driven turbulence has been stressed by evidence of correlation between the dispersion
velocity and star formation rates (Dib et al. 2006; Lehnert et al. 2009). For a convenience, we list
the equations here. With the mass dropout, the mass conservation equation
reads
\begin{equation}
R\frac{\partial\siggas}{\partial t}
+\frac{\partial}{\partial R}\left(R\vr\siggas\right) +R\sigsfr=0,
\end{equation}
where $\vr$ is the radial velocity of the gas, $\siggas$ is the surface density of gas, and $\sigsfr$
is the surface density of star formation rates. All the parameters $\Sigma$, $\dot{\Sigma}_\star$,
$V_{\rm R}$, $\Omega$, ${\cal G}$, $P_{\rm turb}$, $V_{\rm turb}$, $H$, and $\rho_{\rm g}$ are functions
of radius $R$ unless explanations. We neglect mass injection of stellar winds back to
the gaseous disk or escapes of winds from galaxies. Star formation process removes some angular
momentum of gas, thus the conservation equation of angular momentum is given by
\begin{equation}
R\frac{\partial}{\partial t}\left(\siggas R^2\Omega\right)+
\frac{\partial}{\partial R}\left(R\vr\siggas R^2\Omega\right)-
\frac{1}{2\pi}\frac{\partial{\calg}}{\partial{R}}+R^3\Omega\sigsfr=0,
\end{equation}
where $\calg=-2\pi R^3\nu\siggas\left(d\Omega/dR\right)$ is the viscosity torque, $\nu$ is kinematic
viscosity, and $\Omega$ is the angular velocity. Combining equations (1) and (2), we have
\begin{equation}
\frac{\partial{\siggas}}{\partial{t}}=\frac{1}{2\pi R}\frac{\partial{}}{\partial{R}}
     \left\{\left[\frac{d \left(R^2\Omega\right)}{dR}\right]^{-1}
     \frac{\partial \calg}{\partial R}\right\}-\sigsfr.
\end{equation}
Let $X=R^2\Omega$, equation (3) is converted into a more concise form
\begin{equation}
\frac{\partial\siggas}{\partial t}=\frac{1}{2\pi R}\frac{dX}{dR}
       \frac{\partial^2 \calg}{\partial X^2}-A\siggas^{\gamma},
\end{equation}
where the Kennicutt-Schmidt (KS) law as $\sigsfr=A\siggas^{\gamma}$ is used and $A$ is a constant (Kennicutt
1998)\footnote{Star formation law at high$-z$ could be different from the KS law (Gnedin \& Kravtsov, 2010),
but we ignore the difference for simplicity. The influence of the star formation law on the structure of the
gaseous disk can be generally estimated by equations (15-17) and (19-21). See a brief discussion in \S2.2.6.}.
This equation includes the star formation process which has two major effects: consume the gas and
apply energy to make turbulence by SNexp. Imposing adequate boundary and initial conditions to equation (4),
it will produce the structure of the gaseous disks.

\begin{table*}[t]
{\footnotesize\begin{center}
\caption{Summary of all input parameters in the model}
\begin{tabular}{lll}\hline\hline
Parameter      & Units      & Physical meanings and values\\ \hline
$C_\star$      & Gyr$^{-1}$ & the star formation efficiency ($\sim 0.25$)\\
$E_{\rm SN}$   & ergs       & the kinetic energy of supernova explosion ($\sim 10^{51}$)\\
$f_{\rm SN}$   & $\sunm^{-1}$& the parameter of converting SF rates to SNexp rates\\
$V_c$          & $\kms$     & a characterized velocity of the rotation \\
$R_c$          & kpc        & the character radius with velocity $V_c$\\
$R_0$          & kpc        & the initial radius of the gaseous ring\\
$\Sigma_0$     &$\sunm~{\rm pc}^{-2}$& initial surface density of the ring \\
$M_0$          &$10^{10}\sunm $& mass of the initial gaseous ring \\
$\alpha$       & ... & the viscosity parameter \\
$\gamma$       & ... & the index of the KS law ($\sim 1.4$)\\
$\xi$          &$\sunm^{-1}$ & efficiency of the supernova explosion\\
$\Delta_R$     & kpc & the width of the initial ring \\
$\Delta t$     & Gyr & the interval of mergers ($\sim 0.1$)\\ \hline
$f$            & ... & a ratio of radiation to turbulent pressure in photoionization\\
$n_{\rm H}$    & cm$^{-3}$ & density of ionized gas\\ \hline
\end{tabular}\end{center}}
\end{table*}

To further simplify equation (4), we have to specify the relation between $\siggas$ and $\calg$ through the
viscosity parameter $\nu$. Thermal turbulence (currently thought as origin from magnetohydrodynamic-rotation
instability in Balbus \& Hawley 1998) is assumed to make viscosity for transportation of gas angular momentum
as known as the $\alpha-$prescription in standard disk model (Shakura \& Sunyeav 1973). Obviously, the clouds
in the star forming regions are not hot enough to trigger thermal turbulence. However, the turbulent velocity
of $\sim100~\kms$ much higher than thermal sound speed can efficiently make the turbulent viscosity. This
has been supported by that turbulence
excited by SNexp is strong to power dynamical viscosity and transport the angular momentum (Wada
\& Norman 2001; Chen et al. 2009; Paper I and Lehnert et al. 2009). We modify the form of
$\alpha-$ viscosity prescription via replacing the sound speed by the turbulent velocity driven by SNexp
\begin{equation}
\nu=\alpha V_{\rm tur}H,
\end{equation}
where $\alpha$ is a constant, $V_{\rm tur}$ is the turbulence velocity driven by SNexp and $H$
is the thickness of the disk. This prescription can be explained by that clouds are communicating with each
other through the turbulent velocity, and the maximum scale of turbulence can not exceed the height of the
disk. All the uncertainties of the turbulent viscosity are absorbed into the parameter $\alpha$. It is thus
expected $\alpha\le 1$ in the turbulent disk. We take $\alpha$ as a constant throughout the paper.
Detailed transportation of angular momentum driven by SNexp can be found in R\'o\.zyczka et al. (1995)
and briefly in Collin \& Zahn (2008), but we use the simplified version.

Assuming the kinetic energy is channeled into turbulence, energy equation of turbulence excited by
SNexp is given by
\begin{equation}
\frac{\rhog V_{\rm tur}^2}{t_{\rm dis}}=\frac{\rhog V_{\rm tur}^3}{H}=\epsilon {\dot S}_{\star}E_{\rm SN},
\end{equation}
where $t_{\rm dis}=H/V_{\rm tur}$ is the dissipation timescale of the turbulence, $\epsilon$ is the efficiency
converting kinetic energy of SNexp into turbulence, $\dot{S}_\star$ is the SNexp rate and $E_{\rm SN}$ is
the SNexp energy. SNexp rate strongly depends on the initial mass functions [IMF; $N(M_*)$], we have a factor
$f_{\rm SN}=\int_{M_c}^{M_{\rm max}}N(M_*)dM_*/M_{\rm tot}$ to produce SNexp in a time,
where $M_{\rm tot}=\int_1^{M_{\rm max}}N(M_*)dM_*$. We then have $\dot{S}_{\star}=f_{\rm SN}\sigsfr/H$. It is
convenient to introduce $\xi=\epsilon f_{\rm SN}$ for discussion (see \S2.2.5 for details). Considering
relations of $\siggas=\rhog H$ and the KS law, we have
\begin{equation}
\frac{V_{\rm tur}^3}{H}=\xi AE_{\rm SN}\siggas^{\gamma-1}.
\end{equation}
We note that this equation energizing the turbulence is highly simplified, especially the efficiency
$\xi$. The dependence of $\xi$ on the IMF is discussed in \S2.2.5 and \S5. Additionally, SNexp
usually lasts $\sim $ a few $10^4$ years (depending on its surroundings) divided into 3 phases, which
is much shorter than the typical timescale
of dynamical evolution ($\sim 10^{7-8}$ yrs). We thus neglect the time-dependent effects of the SNexp.

We assume an equilibrium in vertical direction
\begin{equation}
P_{\rm tur}=P_{\rm grav},
\end{equation}
where $P_{\rm tur}=\rho_{\rm g}V_{\rm tur}^2$ is the turbulence pressure and $P_{\rm grav}$ is the gravity
in vertical direction. There are two cases to balance the turbulence pressure (Binnery \& Tremaine
2008; Vollmer \& Beckert 2002): A) local stars provide the vertical force and B) central mass dominates.
The two cases have different vertical equilibrium equations. In Case A, $\rhog\ll\rho_\star$ and
$M_{\mathrm g}(R)\ll M_\star(R)$, where $\rhog$ (or $\rho_\star$) is gas (or stellar) density and
$M_{\mathrm g}(R)$ [or $M_\star(R)$] is enclosed mass of gas (or stars) within a radius $R$, namely,
surface density of existed stars dominates. In such a case, we have $P_{\rm grav}=\pi G\Sigma_\star\siggas$
and  $V_{\rm rot}^2=\pi G \Sigma_\star R$ (Binney \& Tremaine 2008, Vollmer \& Beckert 2002),
\begin{equation}
\left(\frac{V_{\rm tur}}{V_{\rm rot}}\right)^2=\frac{H}{R},
\end{equation}
where $V_{\rm rot}$ is the rotation velocity at radius $R$ and $V_{\rm tur}$ is the turbulent velocity.
Such a relation has been suggested by Genzel et al. (2008) to describe a large thin disk for their SINS
galaxy sample. In Case B, $M_{\mathrm g}(R)\leqslant 0.5(H/R)M(R)$, the disk is dominated by the central
mass within
$R$. Then it holds $P_{\rm grav}=\rhog\Omega^2H^2$ (Pringle 1981, Vollmer \& Beckert 2002). Combining
$V_{\rm rot}^2=\Omega^2R^2$ and $\siggas=\rhog H$, equation (8) is rewritten by
\begin{equation}
\frac{V_{\rm tur}}{V_{\rm rot}}=\frac{H}{R},
\end{equation}
which is used to describe a thick or compact disk (Genzel et al. 2008). The different
equilibrium in vertical direction will lead to different structures of the gaseous disk globally.
It is worth pointing out that the vertical equilibrium is sensitive to $\xi$ as shown in subsequent
sections.

We specify rotation curves of the gaseous disk in order to solve equation (4). Following Binney \&
Tremaine (2008), we assume that galaxies have a logarithmic potential leading to
the following rotation curve
\begin{equation}
V_{\rm rot}=\frac{V_c}{\sqrt{1+\left(R_c/R\right)^2}},
\end{equation}
where $R_{\rm c}$ is a critical radius and corresponds to the velocity $V_{\rm rot}=V_{\rm c}/\sqrt{2}$,
and $V_c$ is the velocity at infinity. The rotation velocity and its gradient are then given by
\begin{equation}
\Omega=\frac{V_{\rm rot}}{R}=\frac{V_c}{\sqrt{R^2+R_c^2}}; ~~~~~
\frac{d\Omega}{dR}=-\frac{V_cR}{\left(R^2+R_c^2\right)^{3/2}}.
\end{equation}
We would like to point out that the rotation curves are very similar for the mid-plane gaseous for
both the disk-dominated and bulge-dominated. $R_c$ is a critical radius within which bulge's potential
dominates, and it thus approximates to the size of its bulge. The present model focuses on the dynamics
of the gas at the mid-plane, we thus adopt the same rotation curves for the two cases for simplicity.

{\centering
\figurenum{1}
\includegraphics[angle=-90,scale=0.4]{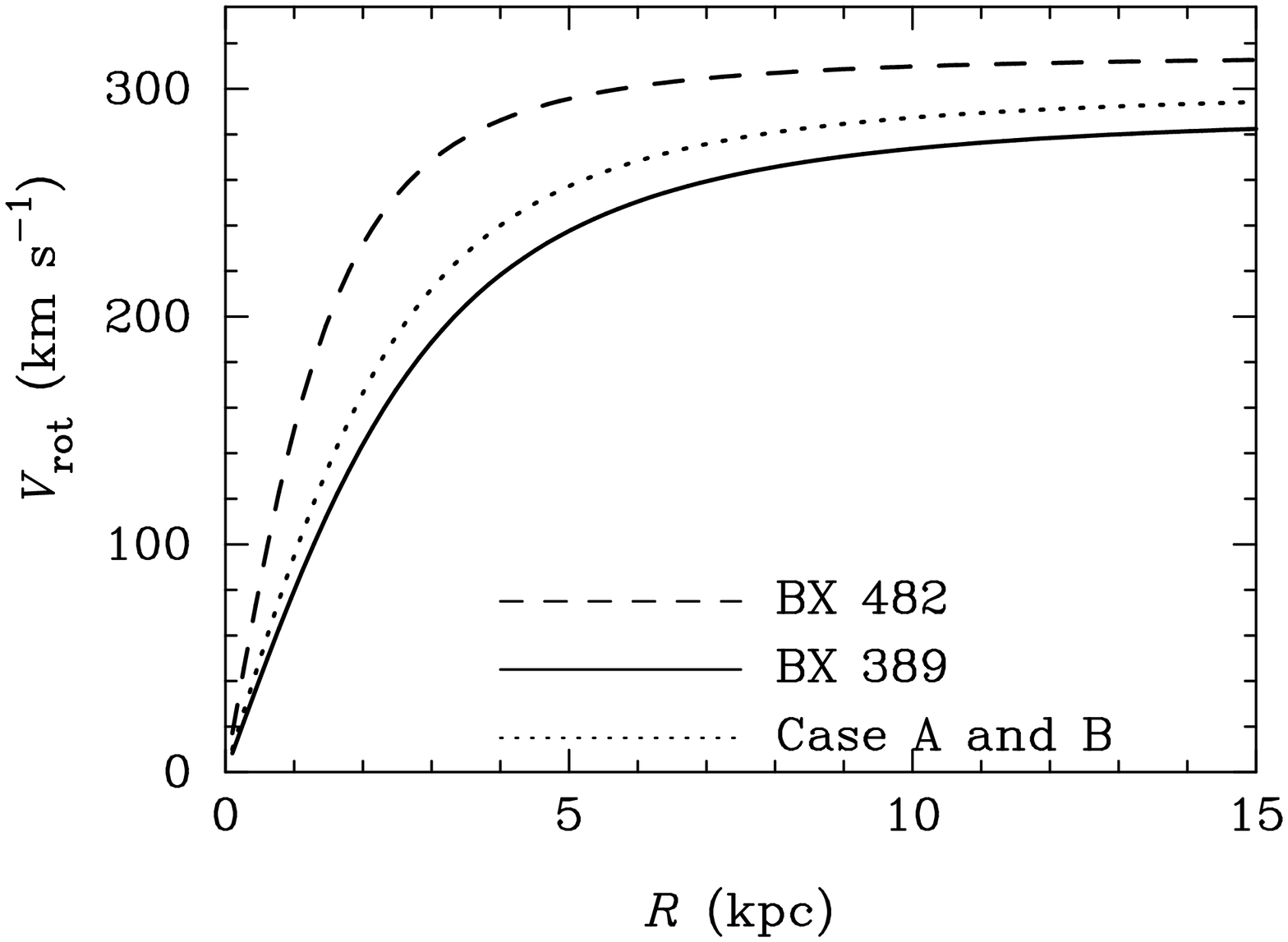}
\figcaption{\footnotesize
Rotation curves for different $V_c$ and $R_c$. Such a rotation curve is very similar to the observed in high$-z$
galaxies (Genzel et al. 2008). Here we plot the curves for BX 389 and BX 482. The dotted line is the rotation
curve used in the calculations for a general case, where we use $V_c=300\kms$ and $R_c=3$kpc.
}
\label{fig1}
}

We list all the equations (4, 5, 7, 8, 9, 10) describing the evolution and structures of the gaseous
disk with feedback of star formation. We stress that the above equations are actually averaged in
vertical direction and the parameters obtained in these equations are the values at the mid-plane
of the disk. Calculations of vertical structures of the disk are definitely important to
compare with observations, but it is beyond the scope of the present paper.

Initial and boundary conditions should be imposed to equation (4) for the disk. The simple
initial condition is a pulsar injection with a Gaussian profile\footnote{Actually, it allows us to
take any form
of the initial profile of the gaseous ring in the numerical calculations. The Gaussian profile as
initial condition is oversimplified, however, this form makes evolutionary properties of the gaseous
ring more clear. For more extended initial gaseous rings, their evolution is approximately
composed of multiple Gaussian rings. The behaviors can be found in Wang et al. (2009).}
\begin{equation}
\siggas(R,0)=\frac{\Sigma_0R_0}{\sqrt{2\pi}\Delta_R}\exp\left[-\frac{(R-R_0)^2}{2\Delta_R^2}\right],
\end{equation}
at radius $R_0$ and with a width $\Delta_R$ as a consequence of a minor merger (Hernquist \& Mihos 1995),
or cold flow (Dekel et al. 2009). Though it is very simple, numerical simulations of Hernquist \& Mihos
(1995) and Dekel et al. (2009) show the reality of the initial condition (equation 13), which is also supported
by high$-z$ galaxies (Genzel et al. 2008). Integrating the ring, we have its total mass
$M_0=\int2\pi \siggas RdR$, which is the merged gas mass from the merged galaxy. The Gaussian
profile used here is not physically necessary, but only for simplification to produce the main features
of the gaseous disk. Any form of initial conditions can be performed by the numerical scheme in the Appendix.

The inner and outer boundary conditions are given by
\begin{equation}
\calg(0,t)=0,~~~~~{\rm and}~~\left.\frac{\partial{\calg}(X,t)}{\partial{X}}\right|_{X=X_{\rm max}}
         =\dot{M}(X_{\rm max})=0,
\end{equation}
where $X_{\rm max}$ is the maximum of the parameter $X$. The inner boundary  is given
by the torque-free condition whereas there is no steady injection to galaxies at the outer boundary.

The viscosity parameter ($\alpha$) is a free one, but it is still constrained by the simple argument
below. The characterized length of medium influenced by SNexp can be estimated by the fact that the
SNexp stops sweeping the medium until its kinetic energy is exhausted by the medium. Since the
kinetic energy of SNexp is equal to the thermal
energy of swept medium, we have $E_{\rm SN}\sim 4\pi R_{\rm SN}^3nm_p/3$, namely,
$R_{\rm SN}\sim 0.2\left(E_{51}/n_2T_2\right)^{1/3}$kpc, where $n_2=n/10^2~{\rm cm^{-3}}$ is the number
density of the medium, $T_2=T/10^2~{\rm K}$ is the temperature and $E_{51}=E_{\rm SN}/10^{51}{\rm erg}$.
This length is comparable with the turbulence size due to SNexp though the actual situation is very
complicated (e.g. Vollmer \& Beckert 2002).  For a thick disk with height of
$\sim 1$kpc, we have $\alpha\sim R_{\rm SN}/H\sim 0.2$, which absorbs the uncertainties of viscosity,
dissipation length and turbulence driven by SNexp. We use $\alpha=0.2$ throughout the present paper.
This is also important for calculations of photoionization, where we find $R_{\rm SN}$ is naturally
about the size of clumps (see equation 32).

In this paper, we use the empirical KS
law $\sigsfr=A\siggas^{\gamma}$, where $\gamma=1.4$,
$A=2.5 \times 10^{-4}$, $\sigsfr$ and $\siggas$ are in units of $\sunm \rm{yr}^{-1} \rm{kpc}^{-2}$ and
$\sunm \rm{pc}^{-2}$, respectively. For simplicity in making equation (4) dimensionless, using
$C_\star=(10^{-6}A) \rm{yr}^{-1}$, we rewrite the KS law as $\sigsfr=C_\star\siggas \siggass^{\gamma-1}$,
where $\siggass=\siggas/\sunm \rm{pc}^{-2}$, for the numerical scheme.

In a summary, we list all the necessary equations and conditions in this subsection. All the parameters
invoked in the models are given in Table 1. We fix several parameters, such as $C_*$, $E_{\rm SN}$,
$\alpha$, $\Delta t$ and $\gamma$ throughout the paper. We do not study the cosmic evolution of galaxy
in the present paper, but focus on the evolution at dynamical timescale. The rotation curves are then
fixed in the evolution. We stress the differences of the present equations from that in Paper
I, in which linear star formation law and constant rotation curves are used for analytical solutions.
Only numerical solutions
exit for the present cases. We describe the numerical scheme to solve the equations in the Appendix.

\figurenum{2}
\begin{figure*}
\centerline{\includegraphics[angle=-90,scale=0.7]{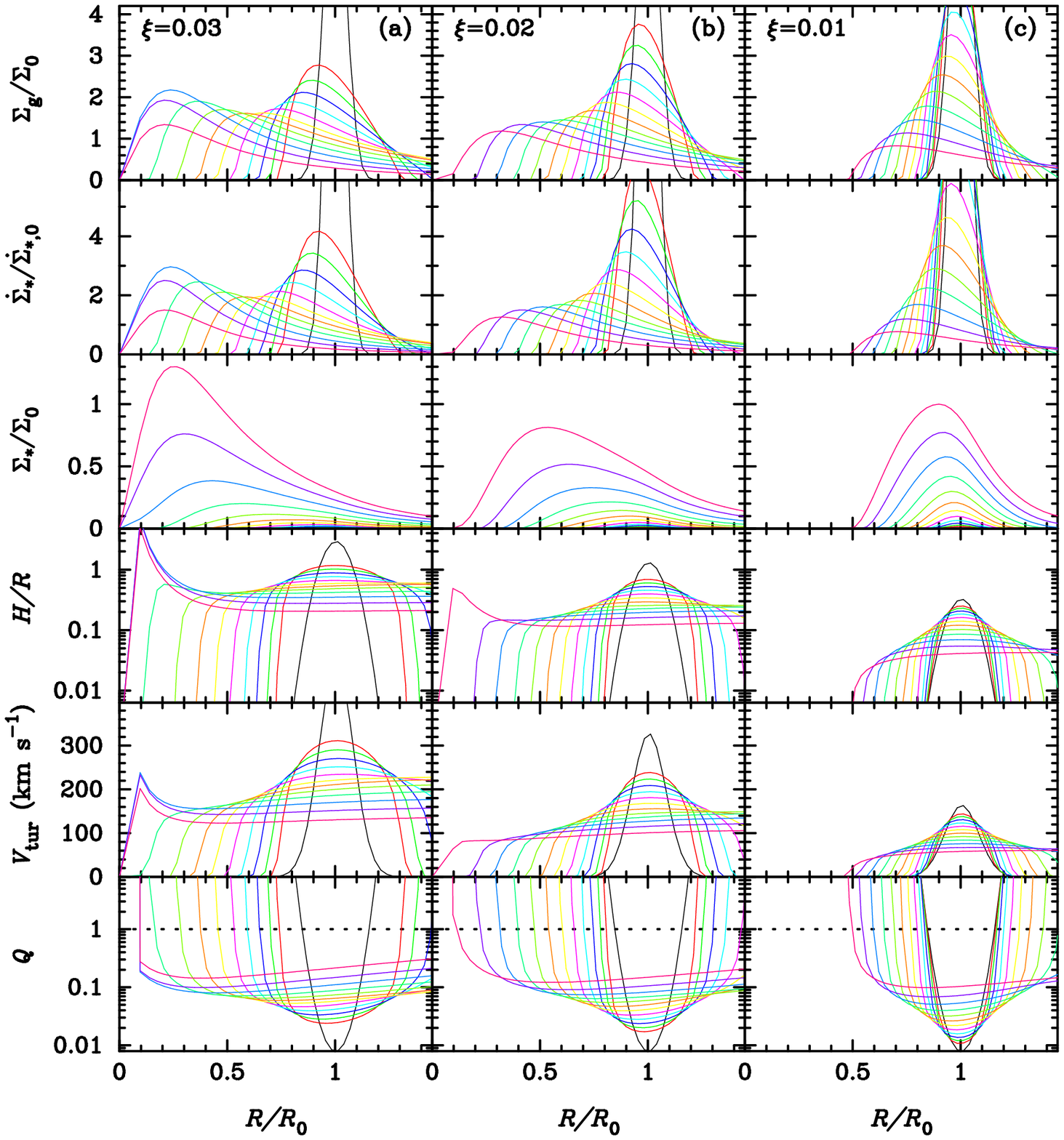}}
\figcaption{\footnotesize Structures and evolution of gaseous disks for Case A. From the upper to
bottom panels are gas surface density, star formation rate surface density, stellar surface density,
height, turbulent velocity and the Toomre parameter, respectively. This plot is for the given initial
conditions with $\Sigma_0=50\sunm {\rm pc}^{-2}$, (corresponding to
$\dot{\Sigma}_{*,0}=0.06\sunm {\rm kpc}^{-2} {\rm yr}^{-1}$), $R_0=10$kpc, $\Delta_R=0.5$kpc,
$M_0=3.25\times 10^{10}\sunm$. Lines with different colors (from  red, green, blue, ..., to
pink) refer to different times from $10^6\rm{yr}$ to $10^9\rm{yr}$ with an interval $\Delta \log t=0.25$.
And the black one is refer to the initial condition.}
\end{figure*}
\label{fig2}
\vglue 0.3cm

\subsubsection{Case A: Stellar mass dominated disk}
For a fast glance of dependence of the solution on the parameters, we have form solutions of the gaseous
disk from combining equations (7) with (9) for rough estimates of the structure of the gaseous disk before
numerically solving the equations. They are
\begin{equation}
V_{\rm tur}=113.8~ \xi_{-2}E_{51}C_{-10}\Sigma_{2}^{\gamma-1}R_{10}V_{300}^{-2}~ {\rm km~s^{-1}},
\end{equation}
\begin{equation}
H=1.44~\xi_{-2}^2E_{51}^2C_{-10}^2\Sigma_{2}^{2\gamma-2}R_{10}^3V_{300}^{-6}~ {\rm kpc},
\end{equation}
\begin{equation}
\nu=33.4~\alpha_{0.2}\xi_{-2}^3E_{51}^3C_{-10}^3\Sigma_{2}^{3\gamma-3}R_{10}^4V_{300}^{-8}~{\rm kpc^2~ Gyr^{-1}},
\end{equation}
where $\alpha_{0.2}=\alpha/0.2$, $R_{10}=R/10~{\rm kpc}$, $E_{51}=E_{\rm SN}/10^{51}{\rm erg}$,
	    $\xi_{-2}=\xi/10^{-2}\sunm^{-1}$, $C_{-10}=C_\star/2.5\times 10^{-10}~{\rm yr^{-1}}$,
            $V_{300}=V_{\rm rot}/300{\rm km s^{-1}}$, $\Sigma_2=\siggas/200\sunm \rm{pc}^{-2}$.

We find from the form solutions that: 1) the velocity dispersion excited by SNexp can be comparable
with the observations; 2) the disk is thin as $H/R\sim 0.1$; 3) the infalling (or advection) timescale
of gas is of $t_{\rm adv}\sim R^2/\nu=R_{10}^2/\nu_{33}\sim 3.0$ Gyr due to the viscosity, where
$\nu_{33}=\nu/33{\rm kpc^2~Gyr^{-1}}$; 4) $V_{\rm tur}$, $H$
and $\nu$ are very sensitive to the rotation velocity $V_{\rm rot}$, which fully depends on the host
galaxies. Clearly the evolution of gaseous disk is sensitive to the host galaxies.

With the rotation curve given by equation (11), we get
$\nu=\zeta_{\rm A}\siggass^{3\gamma-3}\left(R^2+R_c^2\right)^4/R^4$ and
$\calg=2\pi\zeta_{\rm A}V_c\siggass^{3\gamma-3}\siggas (R^2+R_c^2)^{5/2}$, where
$\zeta_{\rm A}=\alpha(\xi E_{\rm SN}C_\star)^3/V_c^8$.
We make the equations dimensionless through $\tau=tC_\star$, $r=R/R_\star$, $r_c=R_c/R_\star$,
$x=X/(V_cR_\star)$, where $R_\star=R_{\rm max}^2 /\sqrt{R_c^2+R_{\rm max}^2}$ making $x_{\rm max}=1$, and
$Y_{\rm A}=\calg/(2 \pi \zeta_A R_{\star}^{5} V_c\sunm{\rm pc}^{-2})=\siggass^{3\gamma-2}(r^2+r_c^2)^{5/2}$.
We rewrite equation (4) as
\begin{equation}
\frac{\partial^2Y_{\rm A}}{\partial x^2}=k_1^{\rm A}(x,Y_{\rm A})\frac{\partial{Y_{\rm A}}}{\partial{t}}+
k_2^{\rm A}(x,Y_{\rm A}),
\end{equation}
where the coefficients are
$$
k_1^{\rm A}(x,Y_{\rm A})=%r\frac{dr}{dx}\phi_A^{\frac{1}{2-3\gamma}}U^{\frac{3-3\gamma}{3\gamma-2}},~~~~~
\frac{C_\star(r_c^2+r^2)^{\frac{9\gamma-11}{6\gamma-4}}}{\zeta_AR_\star^2(3\gamma-2)(2r_c^2+r^2)}
Y_{\rm A}^{\frac{3-3\gamma}{3\gamma-2}},~~~~~
$$
and
$$
k_2^{\rm A}(x,Y_{\rm A})=%r\frac{dr}{dx}\phi_A^{\frac{\gamma}{2-3\gamma}}U^{\frac{\gamma}{3\gamma-2}},
\frac{C_\star(r_c^2+r^2)^{\frac{2\gamma-3}{3\gamma-2}}}{\zeta_AR_\star^2(2r_c^2+r^2)}
Y_{\rm A}^{\frac{\gamma}{3\gamma-2}}.
$$
The initial condition (equation 13) can be converted into
$Y_{\rm A}(x,0)=\left(r^2+r_c^2\right)^{5/2}\siggass^{3\gamma-2}(r,0)$,
where ${\siggass}(r,0)=\Sigma_{0,1}r_0/\sqrt{2\pi}\sigma \exp\left[(r-r_0)^2/2\sigma^2\right]$.
The boundary conditions (equation 14) are converted into $Y_{\rm A}(0,t)=0$ and
$\partial{Y_{\rm A}}/\partial{x}(x_{\rm max},t)=0$. Equation (18) is a non-linear 2nd partial differential
equation. Numerical scheme for the equation is given in the Appendix.

\figurenum{3}
\begin{figure*}
\centerline{\includegraphics[angle=-90,scale=0.7]{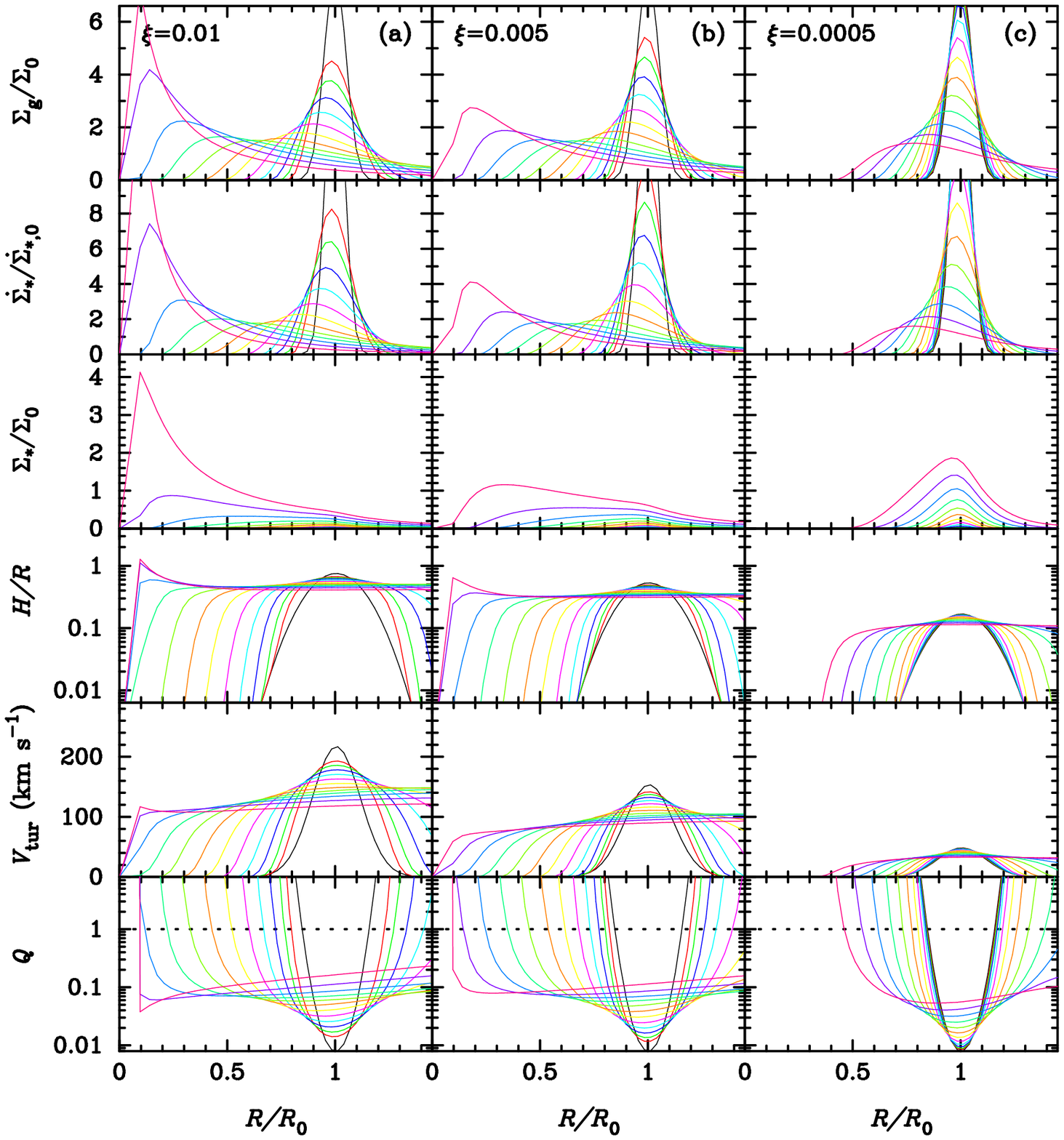}}
\figcaption{\footnotesize Structures and evolution of gaseous disk for Case B. We use the same values of the
initial conditions as in
Fig. 2 for case A, but the parameter $\xi$ is different. The global properties of the gaseous disk is similar
for Case A and B, but they are more sensitive to $\xi$ for Case B.}
\end{figure*}
\label{fig3}
\vglue 0.3cm

\subsubsection{Case B: Central mass dominated disk}
Combining equations (7) and (10), we have
\begin{equation}
V_{\rm tur}=184.7~\xi_{-2}^{0.5}E_{51}^{0.5}C_{-10}^{0.5}\Sigma_{2}^{0.5\gamma-0.5}R_{10}^{0.5}
             V_{300}^{-0.5}~{\rm km~s^{-1}},
\end{equation}
\begin{equation}
H=6.2~\xi_{-2}^{0.5}E_{51}^{0.5}C_{-10}^{0.5}\Sigma_{2}^{0.5\gamma-0.5}R_{10}^{1.5}V_{300}^{-1.5}~{\rm kpc},
\end{equation}
\begin{equation}
\nu=232.5~\alpha_{0.2}\xi_{-2}E_{51}C_{-10}\Sigma_{2}^{\gamma-1}R_{10}^{2}V_{300}^{-2}~{\rm kpc^2~Gyr^{-1}}.
\end{equation}
Obviously, the structures of the Case B disks are quite different from that of Case A: 1) The disks have higher
velocity dispersion as high as $100~\kms$; 2) a quite thick geometry with $H/R\sim 0.5$; 3) the infalling timescale
is much shorter than that of Case A by a factor of $\sim 7$. Moreover, unlike Case A, the disks are not so
sensitive to the rotation curves, however, it still depends on the rotation curves.

We use the similar parameters to make equation (4) dimensionless. From the rotation curve, we get
$\nu=\zeta_B\siggass^{\gamma-1}(R_c^2+R^2)$ and
$\calg=2\pi\zeta_BV_c\siggass^{\gamma-1}\siggas R^4(R^2+R_c^2)^{-1/2}$,
where $\zeta_B=\alpha\xi E_{\rm SN}C_\star/V_c^2$. Using
$Y_{\rm B}=\calg/(2 \pi \zeta_B R_{\star}^{3} V_c\sunm{\rm pc}^{-2})= r^4(r^2+r_c^2)^{-1/2}\siggass^\gamma$,
we get
\begin{equation}
\frac{\partial^2Y_{\rm B}}{\partial x^2}=k_1^{\rm B}(x,Y_{\rm B})
\frac{\partial{Y_{\rm B}}}{\partial{t}}+k_2^{\rm B}(x,Y_{\rm B}),
\end{equation}
where
$$
k_1^{\rm B}(x,Y_{\rm B})
=\frac{C_\star(r_c^2+r^2)^{\frac{3\gamma+1}{2\gamma}}}{\zeta_B\gamma(2r_c^2+r^2)r^{\frac{4}\gamma}}
Y_{\rm B}^{\frac{1- \gamma}{\gamma}},
$$
and
$$
~~~~~k_2^{\rm B}(x,Y_{\rm B})=\frac{C_\star(r_c^2+r^2)^2}{\zeta_B(2r_c^2+r^2)r^4}Y_{\rm B}.
$$
The initial condition is $Y_{\rm B}(x,0)=\siggass^{\gamma}(r,0)r^4(r^2+r_c^2)^{-1/2}$,
and the boundary conditions are identical with Case A.

\subsection{Structure and evolution of a single gaseous ring}
We tested the present numerical scheme in Appendix
by setting $\gamma=1$ to check if the numerical solutions are identical to the analytical
presented in Paper I. It turns out that the numerical scheme in the Appendix
works very well for the non-linear differential equation.

Figure 1 shows several different rotation curves determined by the two parameters $V_c$ and $R_c$. These
curves are characterized by two parts: 1) the inner part, which has a linear increase with radius; and 2)
the outer, which becomes quite flat. The structure of the outer part (beyond $R_c$) is quite similar to the
analytical solutions, however, in the inner part most of the gas will be exhausted for star formation in
the gaseous disks with high and intermediate viscosity. This region is very important to supply gas to
the central supermassive black holes.

We set typical values of $\Sigma_0$, $R_0$, $R_c$ and $V_c$ to solve equations (18) and (22), respectively.
Though the initial values of the parameters influence the properties of the gaseous disk somehow, the main
driver of the solution is obviously the parameter $\xi$, which represents the feedback strength of star
formation. Additionally, KS law is a non-linear relation and leads to some interesting dependence of the
behavior of the disk on the initial surface density. We focus
on the $\xi-$dependence of the disks for Case A and B and discuss the non-linear effects of the KS law.
Table 2 gives values of $\xi$ used in calculations.

\subsubsection{stellar and gas surface densities}
Generally, there are three types of solutions for Case A and B found in the $\siggas-$panels of Figure 2
and 3. For a disk with low-viscosity, as shown by Figure 2{\em c} and 3{\em c}, the gas ring slowly diffuses
spatially only. This behavior is very similar
to the classical gaseous ring described by Lynden-Bell \& Pringle (1974). It forms stars around the initial
radius appearing as a stellar ring or a stellar disk in the diffuse timescale, which is
longer than that of the star formation. Infalling
gas advected by the turbulence-viscosity driven by SNexp is not powerful enough to transport gas inward.
Here we simply ascribe the low viscosity to the low SNexp rates, which could be caused by a very steep
IMF. In such a case, the fraction of young massive stars are fewer than that of flat one, which determines
the SNexp rates and hence $\xi$. We also note that the gas is not able
to diffuse into $R_c$. The gaseous disk is mainly dynamically controlled by the part with constant rotation.
The solution of the low-viscosity ring is very similar to the analytical one in Paper I.

{
\centering
\figurenum{4}
\includegraphics[angle=-90,scale=0.55]{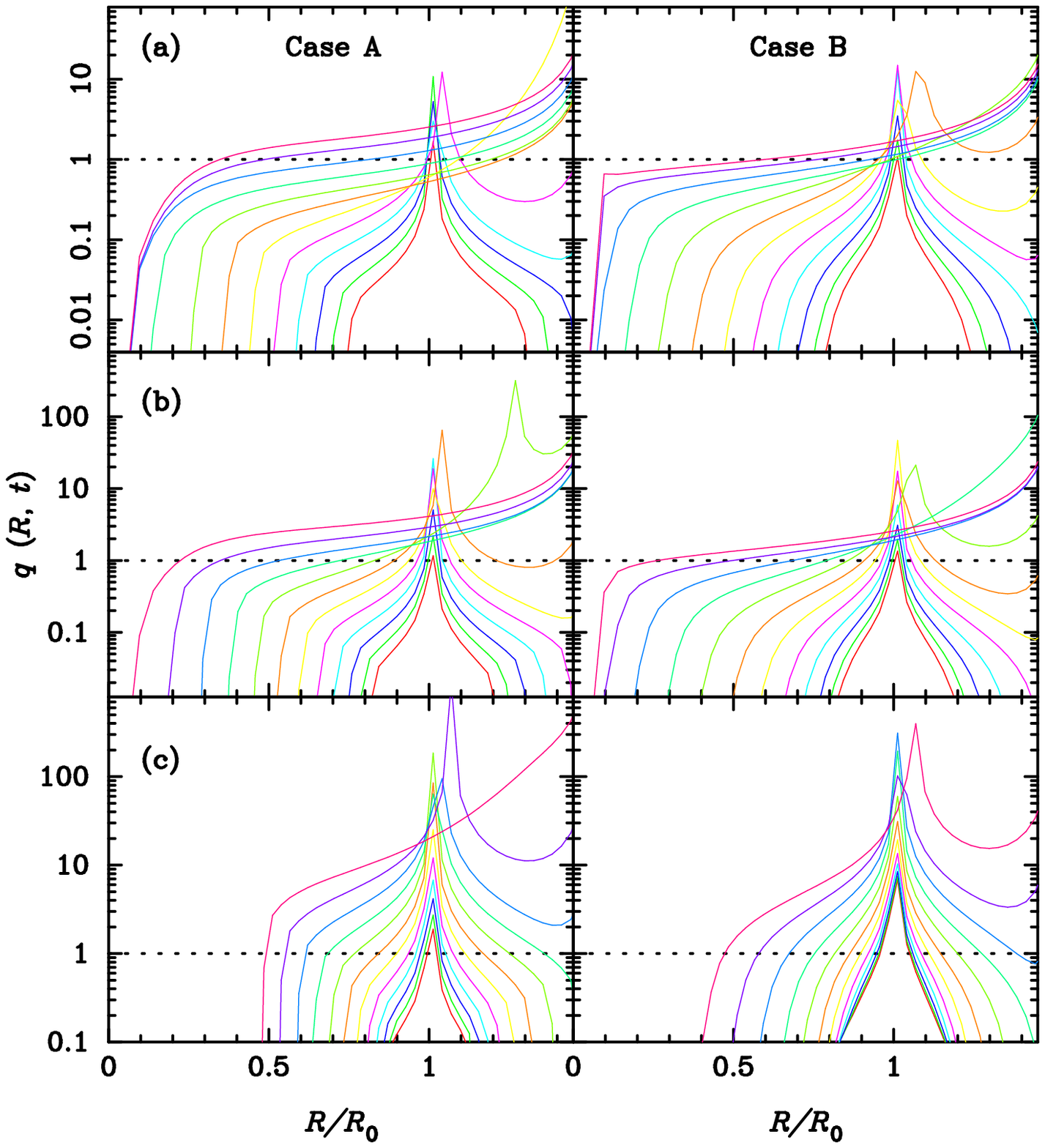}
\figcaption{\footnotesize The ratios of advection to star formation timescales at different radius and
time. $q<1$ means that advection of gas by the SNexp-driven viscosity dominates over the star formation.
The panel {\em a}, {\em b} and {\em c} correspond to that in Figure 2 and 3.}
\label{fig4}
\vglue 0.3cm
}

For a gaseous ring with high-viscosity as shown in Figure 2{\em a} and 3{\em a}, the gas is efficiently
delivered inward by the turbulence-viscosity torque through SNexp, forming a peak and a valley of the
envelope of the $\siggas-$evolution track at $R_{\rm peak}$ and $R_{\rm valley}$, respectively. The
ring has three phases during its evolution. First, it dramatically spreads into a thick
disk within about a few $10^6$yrs. Second the ring keeps a thick disk within a few $10^7$yrs. And finally
it goes into the last phase, in which the disk evolves into a stellar disk within a few $10^8$yr.
Contrary to a ring with low viscosity, the high viscosity ring forms a stellar ring or a stellar disk
far away from the initial radius of the gaseous ring by a distance of $\sim 0.1$ initial radius. The
high-viscosity arises from higher SNexp rates, which imply a high star formation rate or
a top-heavy IMF.

For a gaseous disk with intermediate-viscosity as shown in Figure 2{\em b} and 3{\em b}, star
formation is competing with transportation of gas inward, showing a quite flat distribution of
gas density without peak and valley of the envelope of $\siggas-$evolution track. We note that
the properties of the $\siggas-$evolution track depend on the rotation curves, the initial gas
density and radius, but we focus on understanding the role of star formation in $\siggas-$evolution
track.

The peak and valley of the $\siggas-$evolution track are determined by the roots of the equations
of $\partial \siggas/\partial R=0$ and $\partial \siggas/\partial t=0$. We have $R_{\rm peak}$ and
$R_{\rm valley}$ for the analytical solutions in Paper I. The present properties of $\siggas-$evolution
tracks are similar that in Paper I, but the gaseous disk shows faster evolution than the analytical
solution. This is because the
empirical star formation KS law is steeper than the linear relation. Considering the timescale
of star formation $t_*=\siggas/\sigsfr=\siggass^{1-\gamma}/C_\star=0.6~\Sigma_2^{-0.4}$Gyr,
and gas advection timescale $t_{\rm adv}=R/V_R\approx R^2/\nu=1/\zeta_A R^2\siggass^{3(\gamma-1)}$
for case A and $t_{\rm adv}=1/\zeta_B \siggass^{\gamma-1}$ for case B, the solutions can be qualitatively
understood through the ratio
\begin{equation}
q(R,t)=\frac{t_{\rm adv}}{t_*}=\left\{\begin{array}{l}
6.2~\alpha_{0.2}^{-1}C_{-10}^{-2}\xi_{-2}^{-3}E_{51}^{-3}R_{10}^{-2}V_{300}^8\Sigma_2^{2-2\gamma},\\
            \\
0.9~\alpha_{0.2}^{-1}\xi_{-2}^{-1}E_{51}^{-1}V_{300}^2,\end{array}\right.
\end{equation}
for Case A and B, respectively. It is interesting to find that $q$ just follows $V_{\rm rot}$ for
Case B.

Figure 4 shows the parameter $q(R,t)$. The sharp peaks in these panels correspond to the point
($R_{\rm zero}$) where the drift velocity ($V_{\rm R}$) is very low and close to zero. Beyond $R_{\rm zero}$,
the gas is transported outward whereas within $R_{\rm zero}$ the gas is advected inward. We would like
to point out that only tiny fraction of the gas is transported outward (beyond the initial radius $R_0$)
as shown in the $\siggas-$panels.
It is clear to show from $q-$panels that the low-viscosity leads to $q>1$, namely, star formation dominates
whereas the high-viscosity does $q<1$ with dominant advection of gas.

\subsubsection{geometry and turbulence velocity}
We plot the height of the disks and the turbulence velocity in $H-$ and $V_{\rm tur}-$panel in Figure 2
and 3, respectively. Generally, $H/R\lesssim 1$ holds for Case A and B, and
the disk at the final stage has $H/R\sim 1$.
The gaseous disk becomes very sharp at the inner edge for the cases with high and
intermediate viscosity. The sharp peak of the disk height at its inner edge arises from that the left gas
is bloated by the intensive SNexp during the star formation in light of a relatively high $\xi$.
For a ring with low viscosity, it simply spreads over the space with a constant $H/R$ for all the time.
This is very similar to the classical solution of diffusion problem for a ring with a constant viscosity
(e.g. Lynden-Bell \& Pringle 1974).

We plot the turbulence velocity in Figure 2 and 3. We find that $V_{\rm tur}$ holds a rough constant over
the disk, but evolving into smaller one with time. This is caused by the decreases of star formation rates.
The low viscosity arisen by low SNexp rates only supports a thinner disk. This predicts a relation between
the height and dispersion velocity, however, it is very difficult to test the relation since the height of
the gaseous disks is unknown from observations.

\subsubsection{Self-gravity}
Star formation happens in the self-gravity-dominated regions.
The numerical solutions allow us to test the self-gravity of the gaseous disk through the
Toomre parameter defined as
\begin{equation}
Q=\frac{c_s\kappa}{\pi G\siggas}
\end{equation}
where $\kappa=\sqrt{4\Omega^2+d\Omega^2/d\ln R}$, $c_s$ is the sound speed, we plot $Q$ in $Q-$panel of
Figure 2 and 3.
The parameter $Q$ actually represents the ratio between the vertical gravity to the self-gravity.
$Q<1$ implies a self-gravity dominated solution whereas $Q>1$ means the self-gravity can be neglected.
We find $Q$ is smaller than unity in most regions during the evolution, making the star formation
ongoing. This is in agreement with observations of Genzel et al. (2008) showing evidence for globally
unstable of disk. We would like to stress that the evolution of gaseous disk is controlled by the
SNexp through viscosity, rather than by adjusting the Toomre parameter $Q\sim 1$. This is different
from that in Thompson et al. (2005) who suggest that $Q\sim 1$ keeps by adjusting the star formation
rates. However, the disagreement between Wada \& Norman (2001) and Thompson et al.
(2005) could be removed if the gaseous disk is clumpy, in which the definition of the parameter $Q$
should be modified below.

We note that $Q^{\prime}=V_{\rm tur}\kappa/\pi G\siggas$ is widely used for clumpy disk in Vollmer \&
Beckert (2002) and regarded as a free parameter. Actually $Q'$ is very different from $Q$. Communication
between clumps or clouds is in term of the turbulent speed $V_{\rm tur}$, but the gravitation instability
develops through location sound speed $c_s$. $Q'$ describes the relative strength of the turbulent ram
pressure to the vertical direction self-gravity, and does not deliver the classical meanings of the
Toomre parameter. Unlike Thompson et al. (2005), given the star formation law, $Q'$ can be derived, which
plays an essential role in clumpy disks.

\figurenum{5}
\begin{figure*}[t]
\centerline{\includegraphics[angle=-90,scale=0.6]{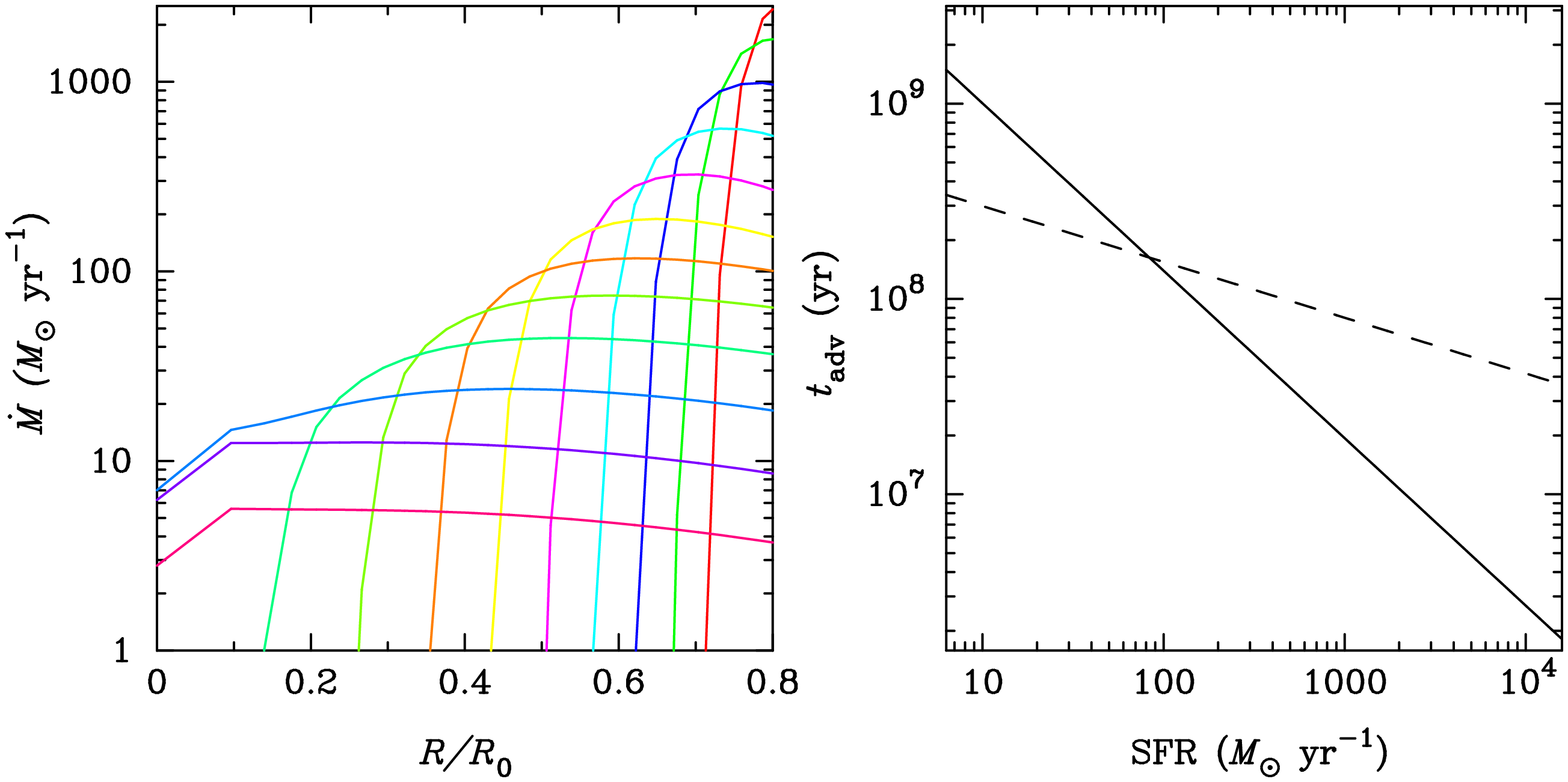}}
\figcaption{\footnotesize {\em Left:} Evolution of accretion rate of gas for Case A, here $\xi=0.03\sunm^{-1}$,
corresponding to fig 2a. Lines with different colors (from red, green, blue, ..., to
pink) refer to different times from $10^6\rm{yr}$ to $10^9\rm{yr}$ with an interval $\Delta \log t=0.25$.
{\em Right:} the timescale of delivering gas with the star formation rates, the solid and dashed lines are
for Case A and B, respectively. We find $t_{\rm adv}\sim 10^8\left({\rm SFR}/10^2\sunmyr\right)^{-0.9}$yr.}
\end{figure*}
\label{fig5}

\subsubsection{The fates of stellar ring}
The secular evolution of the gaseous disk leaves a stellar disk or a stellar ring for high and
low viscosity, respectively. For a gaseous disk with high viscosity, the initial gaseous ring is moved
to inner region through the star formation feedback, and forms a stellar disk with flexible width then,
as shown in the $\Sigma_*-$panel of Figure 2 and 3. It has been suggested by Noguchi (1999) and Elmegreen
et al. (2008) that the bulges in local spiral galaxies form through the dynamical evolution of massive
clumps of stars in high-redshift galaxies. The formation and growth of bulges may be related with fates
of the stellar rings. When $R_{\rm peak}\le R_c$, i.e. the stellar ring is at the outer edge of the
bulges, the fates of the stellar ring might be determined by
friction of the stellar ring with stars in the bulge since they are dense
enough. In such a case,  the stellar friction of the ring with stars in the bulge is driving the
ring to immigrate into the bulge making growth of the bulges. This timescale is determined by
\begin{equation}
t_{\rm imm}\sim 0.3\left(\frac{V_{\rm rot}}{\sigma_0}\right)^2t_{\rm dyn}^*
           \approx 1.5\times 10^6 R_{1}V_{200}^{-1}~{\rm yr},
\end{equation}
where $t_{\rm dyn}^*$ is the stellar dynamical timescale, $R_1=R/1{\rm kpc}$,
$V_{200}=V_{\rm rot}/200\kms$, $\sigma_0$ is the dispersion
velocity (see eq. 6 in Genzel et al. 2008) and $V_{\rm rot}\approx \sigma_0$ is used. We find that the
timescale is significantly shorter than the typical episodic lifetime of star formation, namely, the
ring immigrates into the bulge very fast. It happens for the case with high-viscosity. When the
viscosity is low, the stellar ring will mixture with stars around its initial location. In such a case,
the galactic stellar disk is growing.

\subsubsection{Dependence on $\xi$}
For the linear star formation law, we have the critical value of $q_c$ in Paper I, which determines if there
is a flat envelope of the $\siggas-$evolution track. $q_c$ depends on the details of the index of star
formation law, here we approximate $q_c\sim 1.0$ in light of the analytical solution. From equation (23), we have
the critical value of $\xi$,
\begin{equation}
\xi_c=\left\{\begin{array}{l}
0.018~\alpha_{0.2}^{-1/3}q_c^{-1/3}C_{-10}^{-2/3}E_{51}^{-1}R_{10}^{-2/3}V_{300}^{8/3}\Sigma_2^{(2-2\gamma)/3},\\
            \\
0.009~\alpha_{0.2}^{-1/3}q_c^{-1}E_{51}^{-1}V_{300}^2,\end{array}\right.
\end{equation}
for Case A and B, respectively. We refer the high-viscosity to $\xi>\xi_c$ whereas the low-viscosity to
$\xi<\xi_c$. We find that this simple estimation agrees with the numerical results as shown in Figure 2
and 3 for the cases of the high and low viscosity.

It has been firmly indicated from observations that galaxies are undergoing episodic star formation
(e.g. Glazebrook et al. 1999). Kinetic feedback from SNexp in star formation depends on episodic active time
($\Delta t_{\rm G}$). For an episodic activity of star formation, the minimum stellar mass to produce an
SNexp within $\Delta t_{\rm G}$ should be
\begin{equation}
M_{\rm c}/\sunm\ge 7.0~\Delta t_{0.1}^{-0.4},
\end{equation}
where we use the lifetime of hydrogen main sequence stars as $t_{\rm MS}=13\left(M_*/\sunm\right)^{-2.5}$Gyr
and $\Delta t_{0.1}=\Delta t_{\rm G}/0.1{\rm Gyr}$. Otherwise stars less than this critical one will not play
a role in the turbulent viscosity in star forming galaxies. Solution of the dynamical structure sensitively
depends on the parameter $\xi$, which is
\begin{equation}
\xi=\epsilon f_{\rm SN} \approx \epsilon\frac{M_{\rm min}^{\beta-2}}{M_c^{\beta-1}}
    \approx 1.0\times 10^{-2}\epsilon_1m_{0.1}^{0.35}m_7^{-1.35}M_{\odot}^{-1},
\end{equation}
where $m_{0.1}=M_{\rm min}/0.1\sunm$, $m_7=M_c/7\sunm$ and the Salpeter
mass function with $\beta=2.35$ is used. Here we show how the viscosity ($\xi$) depends on the IMF. Though
$\xi\propto \Delta t_{0.1}^{0.54}$ for $7\sunm$ stars, we show $\xi\propto \Delta t_{0.1}^{3.1}$ for more
massive stars in \S5. Clearly, the kinetic feedback strongly depends on the IMF for
high$-z$ galaxies.

We would like to point out that $f_{\rm SN}$ can be in a quite large range in light of the IMF. For a
top-heavy IMF, $M_{\rm min}\approx 0.5(1+z)^2\sunm$, where $z$ is redshifts (Dave 2008), we have
$M_{\rm min}=8\sunm$ for stars at $z=3$ and $f_{\rm SN}\approx 4.6\times 10^{-2}\sunm^{-1}$. On the other hand,
for a flat IMF, $\beta\approx 1.0$ (Kroupa 2001), we have $f_{\rm SN}=0.125\sunm^{-1}$. The $f_{\rm SN}$
range is then expected to $0.01-0.1\sunm^{-1}$. For a given $\xi$, the larger $f_{\rm SN}$, the smaller
$\epsilon$. The exact values of $\epsilon$ and $f_{\rm SN}$ are not important, but the $\xi$ with the
similar range of $f_{\rm SN}$.

\subsubsection{Non-linear effects of star formation rates}
Compared with results in Wang et al. (2010), the non-linear star formation law used in the present paper
generally leads to a dependence of the viscosity  on gas density (see equations 17 and 21).
%a higher $\dot{S}_*$ for a given gas density than that of the linear law.
This non-linear effect results in
a dependence of the structure and evolution of the gaseous disk on the initial conditions, especially on
the initial surface density of the ring.
For a linear star formation law, we find that the critical value of $\xi$ is
$\xi_c\approx 0.5C_{-8}/R_{10}^2$ significantly larger than $\xi_c$ (equation 26), where $C_{-8}=C_*/10^{-8}$
and $R_{10}=R_0/10{\rm kpc}$ is the initial
radius of the ring in Paper I, and is independent of the initial gas mass and density.
For the KS law, $\gamma=1.4$ leads to the dependence of $q(R,t)$ on the surface density as shown by equation
(23). Calculations are made for several cases with different initial surface density. We find that
the gas surface density and stellar distribution for a non-linear star formation law are generally steeper
than that of linear one, and the larger $\gamma$, the steeper $\siggas$ and stellar distribution are.
We do not provide the relevant figures here for saving pages.

In a brief summary of this section, we obtain the numerical solutions of the gaseous disk and their properties.
We draw a conclusion that a gaseous ring is going to form a stellar disk far away from it initial location
through turbulent viscosity driven by SNexp.

\subsection{Delivering gas to galactic centers}
With the structure of the gaseous disks, we obtained inflow rates for the issue as to feed SMBHs located at
galactic centers. Figure 5 shows the inflow rates as a function of radius and time. We find that the resultant
inflow rates are at a level of  a few $\sunmyr$.
A vanishing boundary condition for numerical convenience is used in this model at inner edge of the gaseous disk,
the present model does not produce the true "central region". However,
the model is enable to provide the timescale of delivering gas and its dependence on the SFR.
From \S2.1.1, \S2.1.2 and \S2.2.1, we have the timescale $t_{\rm avd}=R/V_R$,
\begin{equation}
t_{\rm adv}=\left\{\begin{array}{l}
      1/\zeta_A R^2\siggass^{3(\gamma-1)}=0.1~\alpha_{0.2}^{-1}\xi_{0.03}^{-3}E_{51}^{-3}C_{-10}^{-3}V_{300}^{8}
      \Sigma_{2}^{3(1-\gamma)}R_{10}^{-2}~{\rm Gyr},\\
            \\
      1/\zeta_B \siggass^{\gamma-1}=0.1~\alpha_{0.2}^{-1}\xi_{0.03}^{-1}E_{51}^{-1}C_{-10}^{-1}V_{300}^{2}\Sigma_{2}^{1-\gamma}~
      {\rm Gyr},\end{array}\right.
\end{equation}
where $\xi_{0.03}=\xi/0.03 \sunm^{-1}$ for Case A and B. Here we only consider the high viscosity case, which is able
to deliver gas into the center regions. Additionally, there are many evidence show that the IMF is top heavy in the
circumnuclear star forming region that means high viscosity by SNexp in hundreds parsecs scale (see equation 28).
The simple estimation of the timescale shows $t_{\rm adv}\propto \siggas^{-1.2}$ and $t_{\rm adv}\propto \siggas^{-0.4}$,
where $\gamma=1.4$ is used, for Case A and B, respectively, indicating that the timescale decreases with star formation
rates. For a more clear impression, we use an approximation ${\rm SFR}\sim \pi R^2\sigsfr$ as a global SF rates to
show relation of $t_{\rm adv}$ versus SFR in the right of Figure 5. For typical star forming galaxies with star formation
rates are of a few $10^2\sunmyr$, we find that the advection timescale is less than $10^8$ yr, namely, SMBH activity
appears when starbursts precede $t_{\rm adv}\lesssim 10^8$yr for Case A. Case B shows much flat dependence of the
time lag on the SFR. It should be noted that this difference is caused by the different dependence on $\siggas$.

Transportation of gas in the $\sim 1$pc scale (inside the dusty torus) in AGNs has been discussed by Wang et al. (2010).
They show that the SNexp-driven inflows can form the Shakura-Sunyaev accretion disk around the SMBH with typical
sub-Eddington rates, naturally explain the well-known metallicity-luminosity relation. It is therefore that the
delivering timescale of gas is just the time lag of AGNs after starbursts. We have to stress that the effects of
thermal heating due to SNexp have not been included as feedback to the process of the star formation in the
current model, which surely lower SF rates in the gaseous disk. This should be investigated by including of the
thermal structure of the disk to improve the
current model further.

\subsection{Multiple rings due to successive minor mergers}
The scheme described above is for a single episode of activity, however, recent observational results
suggest that rapid but more continuous gas accretion via "cold flows" (Keres et al. 2005;
Ocvirk et al. 2008; Brooks et al. 2009; Dekel et al. 2009)) and/or minor mergers likely are
playing an important role in driving star formation and mass growth of the massive star-forming galaxies
at $z>1$ (e.g. Noeske et al. 2007a,b; Elbaz et al. 2007; Daddi et al. 2007)\footnote{We would like to
point out that the current model does not rely on sources of
the gas supply whatever it is by minor mergers or cold flows. The present model
does not predict a relation between merger rates and $V_{\rm rot}/V_{\rm tur}$.} . More interestingly,
current observations of high$-z$ galaxies show there is star formation inside of bulge with potential
disk geometry ($<$ a few kpc) while a star forming ring in the outer regions (Genzel et al. 2008).
Successive minor mergers indicated by observations are happening more frequently at high$-z$ galaxies
(Stewart et al. 2009; Genzel et al. 2008;
Fo\"rster Schreiber et al. 2009). When the minor merger timescale (the inverse of minor merger frequency)
is longer than the advection timescale of gaseous disks,
the multiple rings are evolving independently. In such a case, we only need include multiple minor
mergers by an interval $\Delta t$ for high$-z$ galaxies. We do not include the interaction among the
multiple rings.

A bulge+ring structure  has been often found in SINS sample (Genzel et al. 2008). It may be due to
multiple mergers in term of the merger rates with an
interval of $10^8$ yr or so (Lacey \& Cole 1993). The intensive star formation taking place in the bulge
is from the gas transported by the turbulent viscosity driven by SNexp and the ring or thick disk with
star formation is due to a recent merger. For two successive minor mergers, the first took some gas to the
galaxy and made a gaseous ring, then it evolves and is transferred into circumnuclear region in about
$10^{7-8}$yr. The second minor merger will occur and create another ring
structure. As a consequence of multiple minor mergers, the galaxy appears as ring+bulge.

{
\centering
\figurenum{6}
\includegraphics[angle=-90,scale=0.5]{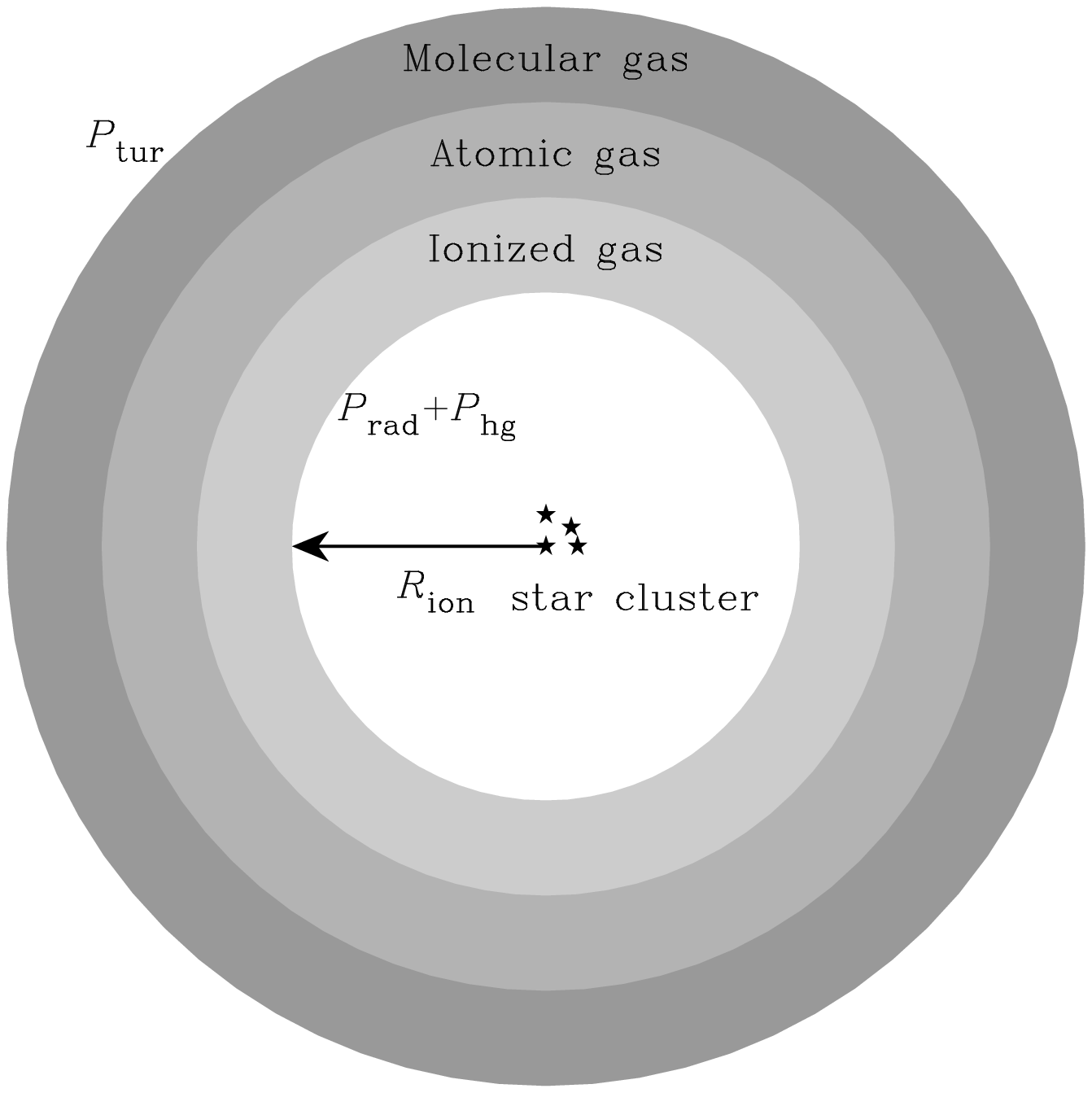}
\figcaption{\footnotesize The scenario of photionization of clumps hosting a star cluster. The
geometry of the clump is enclosed, but the actual should be open like in M17 (their Figure 2
in Pellegrini et al. 2007) in order to transport kinetic energy of the SNexp into turbulence.
The pressure balance is given by $P_{\rm tur}=P_{\rm rad}+P_{\rm hg}$ and
$P_{\rm hg}=\varsigma P_{\rm rad}$, where $\varsigma$ is around unity at the illuminated
face (see text for a detail).}
\label{fig6}
\vglue 0.3cm
}

For different minor mergers, the initial locations and the gas supply from the donor galaxies is different,
leading to different behaviors of evolution and structure of the gaseous disks. As illustrations
of the undergoing physical process in high$-z$ galaxies, we try to apply the present model to two
representative galaxies (BX 389 and BX 482) to show histories of evolution. We find that multiple
mergers can set up the
bulge+ring structure (see Figure 12 and Figure 14).  We would like to stress that the multiple minor mergers
are not arbitrarily adopted. On the contrary, the multiple model has been imposed by more strong constraints:
1) the distance of the two successive
rings, which is driven by the dynamical evolution; 2) the interval of the two successive minor mergers,
testing the merger rates; 3) the relative strength of star formation rates.

\figurenum{7}
\begin{figure*}[t]
\centerline{\includegraphics[angle=-90,scale=0.65]{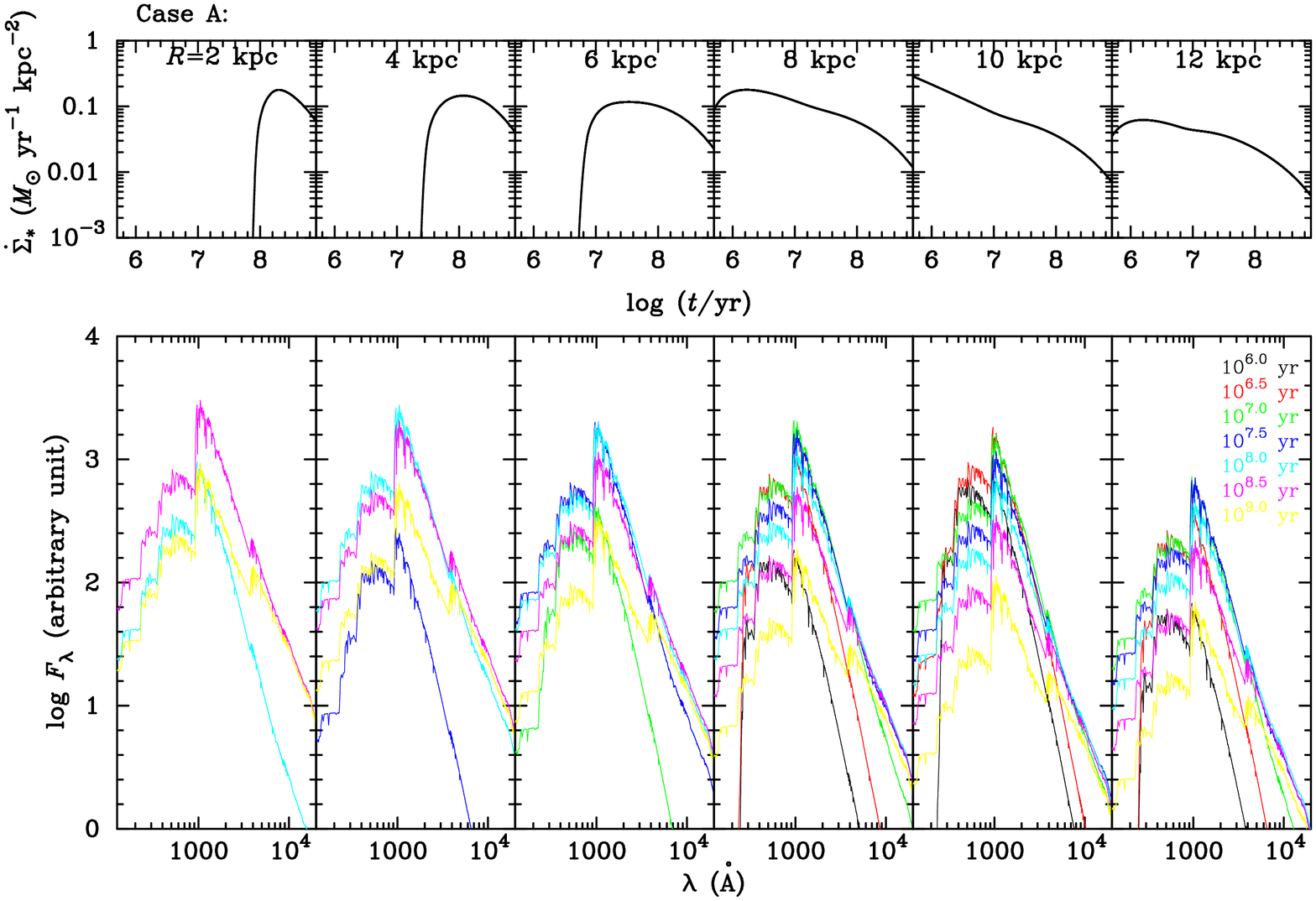}}
\figcaption{\footnotesize The upper panels show star formation histories at different radii for Case A.
The lower panels are the ionizing spectra, which correspond to the star formation rate history in
the upper panel. The initial radius of the gaseous ring is at $R=10$kpc. The gaseous disk
structure used corresponds to Figure 2{\em a}.}
\label{fig7}
\end{figure*}
\vglue 0.3cm

\section{Photoionization of gaseous disks}
With the given structures of the evolving gaseous disk, we are able to, in principle, calculate the
emission from the disk through photoionization by formed stars. For this goal, we have to specify
following parameters: 1) ionizing continuum and luminosity; 2) the distance of ionizing source to
the illuminated face and geometry of photoionization; 3) density of the ionized gas and 4) chemical
composition. Star formation happens and forms a cluster mostly
embedded in clumps.  Each cluster containing a dozen to many millions of stars (Weidner \& Kroupa 2005),
which are surrounded by the layers of the $\HII$ region, the photon-dominated region (PDR) and the molecular
cloud. Generally, clusters are the energy source of ionizing their surrounding gas (e.g. Pellegrini et al.
2007). Photoionization of clouds depends on several parameters, such as, the ionization parameters, geometry,
density of the clouds and temperature. Following Lehnert et al. (2009), we
use the ISM metallicity  for all the radii of the gaseous disks. Figure 6 illustrates
the geometry of clusters and physical conditions
of photoionization in one clump. We assume a closed sphere ionized by the center star cluster.

We use the CLOUDY\footnote{http://ferland.org/cloudy/} to calculate the emission from the gaseous disk.
The photoionization is mainly determined by the ionization parameter defined by
\begin{equation}
\calu(R,t)=\frac{\Phi_{\rm H}(R,t)}{n_{\rm H}(R,t)c},
\end{equation}
where $\Phi_{\rm H}$ (${\rm cm^{-2}s^{-1}}$) is the surface flux of ionizing photons, $n_{\rm H}$ is the
total number density of hydrogen and $c$ is the light speed (Osterbrock \& Ferland 2006). Dopita et al.
(2000) and Kewley et al. (2001) carry out detailed calculations of photoionization of star forming galaxies,
but they took the ionization parameter $\calu$ as a free parameter. Unlike these papers, we calculate the
ionization parameter in term of the given structure of the gaseous disk for photoionization.

\figurenum{8}
\begin{figure*}[t]
\centerline{\includegraphics[angle=-90,scale=0.65]{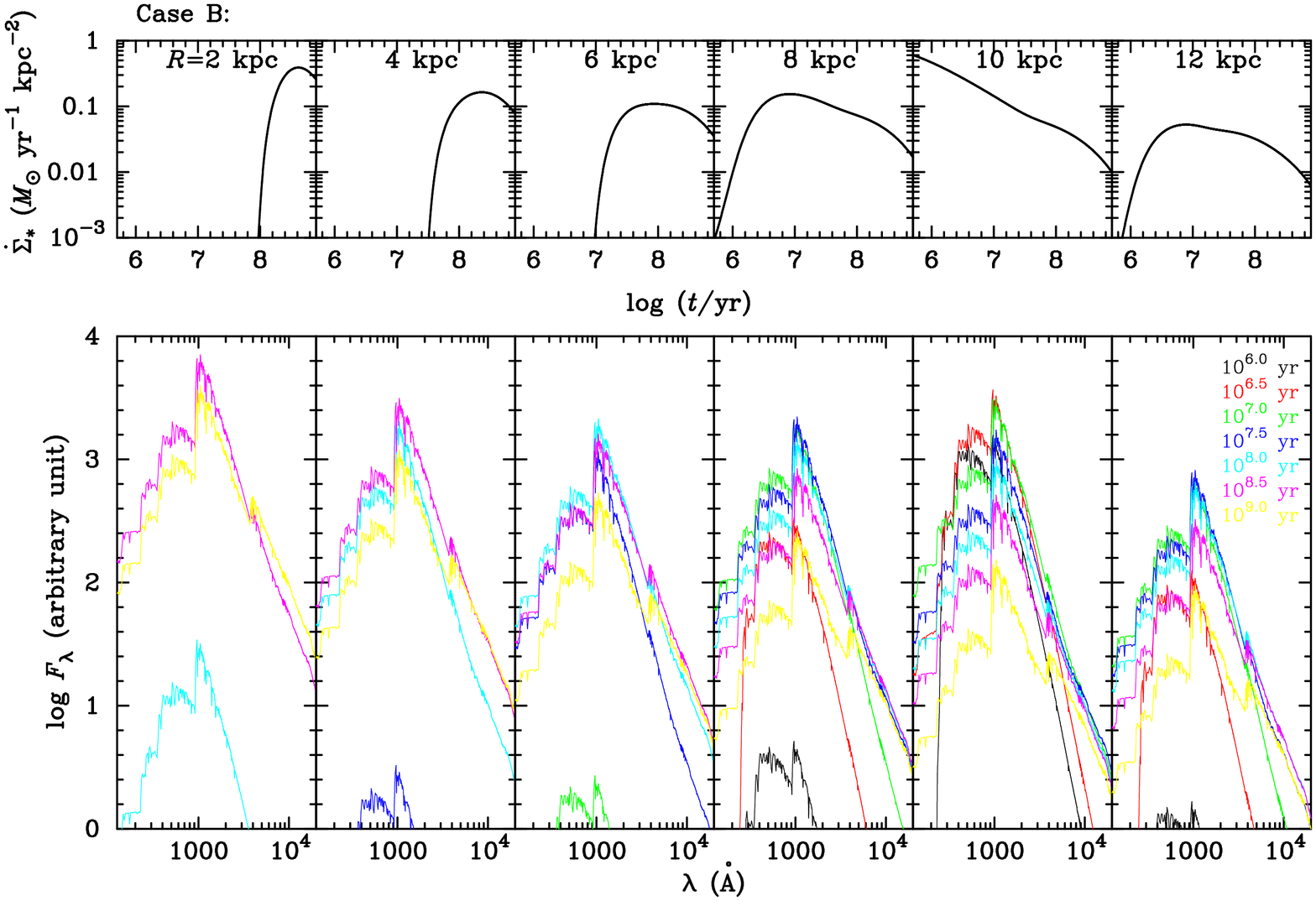}}
\figcaption{\footnotesize  Star formation histories and ionizing spectra as Figure 7, but are for Case B.
The gaseous disk structure used corresponds to Figure 3{\em a}.}
\label{fig8}
\end{figure*}
\vglue 0.3cm

\subsection{Ionizing spectrum}
To calculate the emission from the disk, we divide it into 50 circular belts from 1 to 15 kpc. The
surface gas density and star formation rates are given for each belt by the dynamical equations
numerically. The status of the photoionized gas in each belt is determined by the radiation fields
from the cluster. The ionizing spectrum, in principle, is determined by the IMF, the upper mass
cutoff, cluster age and star formation history, etc. In this paper, we use
the GALAXEV (Bruzual \& Charlot 2003; hereafter BC03) to calculate the ionizing spectra of young
stars in star formation regions as the incident spectra.

Given the initial mass function, metallicities and star formation history, the ionizing photons
from a stellar population is given by
\begin{equation}
\Phi_{\rm H}(R, t)=\frac{1}{4\pi R_{\rm ion}^2}\int_{\nu_c}^\infty \frac{L_{\nu}(R,t)}{h\nu}d\nu,
\end{equation}
where $L_{\nu}(R,t)$ is the total spectral luminosity at time $t$ and radius $R$, and $R_{\rm ion}$
is the radius of the illuminated face of the ionized shell. We should note that $L_{\nu}(R,t)$ relates
with the stellar evolution and is given by  GALAXEV code. In the calculations, we take the minimum star
mass $M_{\rm min}=0.1\sunm$ and the maximum stars $M_{\rm max}=100\sunm$. We use basic SSP models m62
(solar metallicity, Padova 1994 track) and Chabrier IMF for the ionizing source.

\begin{table}
{\footnotesize
\begin{center}
\caption{Values of parameters in calculations}
\begin{tabular}{ccccccc}\hline\hline
\multicolumn{3}{c}{Case A} & &\multicolumn{3}{c}{Case B}\\ \cline{1-3}\cline{5-7}
$\xi(\times 10^{2})$ & $\nu_0$& $q_0$ & &$\xi(\times 10^{2})$& $\nu_0$ & $q_0$ \\ \hline
3.0 & 171.1 &0.70 & & 1.0 & 133.56 &0.90 \\
2.0 & 50.69 &2.36 & & 0.5 & 66.78 &1.79 \\
1.0 & 6.34  &18.87& & 0.05& 6.68 &17.9 \\ \hline
\end{tabular}
\end{center}
}
\end{table}

With the typical values of the parameters given in Table 2,
we calculate the ionizing spectra from the star formation at different radii and times, which are
shown in Figure 7 and 8 for Case A and B at $R=2,4,6,8,10,12$kpc and
$t=10^6, 10^{6.5}, 10^7, 10^{7.5}, 10^8, 10^{8.5}, 10^9$ years, respectively. The parameter $\alpha$
is fixed for each model in this section. The ionizing fluxes
just follow the star formation rates. The Chabrier IMF is used, the spectra thus keep a constant
shape unless at the early stage of star formation and at the stage with a dramatic decrease. The
differences of the ionizing spectra for Case A and B result from the dynamical structure and
the star formation history.

\subsection{ionization of clumps}
We obtain dynamical structure of gaseous disk through continue medium equations, but we have to
connect the averaged structure with clump structures. Following Vollmer \& Beckert (2002), we
assume the turbulence with a spectrum of $E(k)\propto k^{-2}$, where $k$ is the wave number of
the turbulence. The clumps have a characterized size comparable with the turbulent length
($l_c\sim R_{\rm SN}$). We have the characteristic time scale that turbulence crosses the clumps
\begin{equation}
t_{\rm tur}=\frac{l_c}{V_{\rm tur}},
\end{equation}
where $l_c$ is the characterized scale of the clumps, is equal to the gravitational free fall time
$t_{\rm ff}^*=1/\sqrt{G\rho_c}$, where $\rho_c$ is the clump density and $G$ is the gravitational
constant. The characteristic scale of clump is approximated by $l_c=\alpha H$ (Pringle 1981). We
have $\rho_c=n_cm_p=G^{-1}\left(V_{\rm tur}/\alpha H\right)^2$
\begin{equation}
n_c=\left\{\begin{array}{ll}1.5\times 10^3~\alpha_{0.2}^{-2}\xi_{-2}^{-2}C_{-10}^{-2}E_{51}^{-2}R_{10}^{-4}V_{300}^{8}
       \Sigma_2^{2-2\gamma} {\rm cm}^{-3},\\
            \\
0.2\times 10^3~\alpha_{0.2}^{-2}R_{10}^{-2}V_{300}^2 {\rm cm}^{-3},\end{array}\right.
\end{equation}
for Case A and B, respectively. Equation (33) shows $n_c\propto \siggas^{-0.8}$ for Case A, $n_c$
tends to very large for two side edges of the ring since $\siggas$ tends very small there. It is
then expected that the two edges of the $n_c$ could be unphysical.
Clusters within $\HII$ usually have a typical size of 0.1 to 10 pc,
which is much smaller than the height of the gaseous disk at radius $R$.
We introduce the filling factor $\calc$, which is defined by $\calc=\rho_{\rm g}/\rho_{\rm c}$
\begin{equation}
\calc=\left\{\begin{array}{ll} 3.8\times 10^{-3}~\alpha_{0.2}^{2}R_{10}V_{300}^{-2}\Sigma_2,\\
            \\
6.2\times10^{-3}~\alpha_{0.2}^{2}\xi_{-2}^{-0.5}E_{51}^{-0.5}C_{-10}^{-0.5}
\Sigma_{2}^{1.5-0.5\gamma}R_{10}^{-0.5}V_{300}^{-0.5},
\end{array}\right.
\end{equation}
for Case A and B, respectively, where $\rho_{\rm g}=\siggas/H$ is the averaged mass density of gas as in
continue disk and $\rho_{\rm c}$ is the mass density of clumps. We will see $\calc\sim 10^{-3}$ (in \S4)
if the density of clumps is determined by observations of $\SII\lambda 6716,6731$ lines (Lehnert et al.
2009). We note that the averaged ratio of cloud size to distance among clouds is about $\calc^{1/3}\sim 0.1$
which is enough for the validity of the fluid approximation in equation (1).

We further specify the geometry of the photoionized clump as
illustrated by Figure 6. Stellar winds, radiation and supernova produce a cavity of hot gas inside the
clump by pushing the ionization front outward (e.g. Pellegrini et al. 2007; Osterbrock \& Ferland 2006).
We take the wind-nebular interacting surface as the illuminated surface with a radius $R_{\rm ion}$.
If the radiation pressure from stars ($P_{\rm rad}$) is given, we have
\begin{equation}
R_{\rm ion}=\left(\frac{L}{4\pi P_{\rm rad} c}\right)^{1/2},
\end{equation}
where $L$ is the luminosity of the stars in the cluster. In the following, we make an attempt to build
up the physical connection between the photoionization process and the dynamical structure of the gaseous
disk.

The turbulent pressure balances the total internal pressures of the hot gas thermal pressure and radiation
field pressure from stars, namely, $P_{\rm tur}=P_{\rm hg}+P_{\rm rad}$, where $P_{\rm hg}$ is the thermal
pressure of the hot gas. In principle, we have to calculate the hot gas pressure in term of stellar
evolution and winds, however, it is very complicated and beyond the scope of the present paper. For
simplicity, we introduce a parameter $\varsigma$, $P_{\rm hg}=\varsigma P_{\rm rad}$, we have
\begin{equation}
P_{\rm rad}=fP_{\rm tur},
\end{equation}
where $f=1/(1+\varsigma)$, $P_{\rm tur}$ is given by the structure of the gaseous disk. Though the exact
value of $\varsigma$ is not known, it should be around unity $\varsigma\sim 1$, otherwise the
photoionzation is very weak for $\varsigma\gg1$ or the cavity is undergoing formation through the strong
radiation pressure from stars for $\varsigma\ll 1$. We then we have $f\sim 0.5$. The photoionization is
linked with the dynamical structure of the gaseous disk in light of equation (36). We have the ionization
parameter
\begin{equation}
\calu=\frac{F_{\rm ion}}{h\langle\nu\rangle n_{\rm H}c}
     \approx\frac{P_{\rm rad}}{h\langle\nu\rangle n_{\rm H}}
     =f \frac{m_pc^2}{h\langle\nu\rangle}\frac{P_{\rm tur}}{\rho_{\rm c}c^2},
\end{equation}
where $F_{\rm ion}=L_{\rm ion}/4\pi R_{\rm ion}$ is the ionizing flux, $L_{\rm ion}$ is the ionizing
luminosity, $m_p$ is the proton mass, $h$ is the Planck constant, $\rho_c=n_{\rm H}m_p$ and
$\langle\nu\rangle$ is the average frequency of ionizing photons. The approximation is only used to
clarify the dependence of $\calu$ on $f$, rather than used in calculations. $F_{\rm ion}$ is obtained
from the stellar spectra in calculations. We use a constant density of gas throughout the
clump. From the illuminated face, the ionization structure of the shell can be self-consistently
determined by heating and cooling in CLOUDY for the given $\calu$ at the illuminated surface.

{
\centering
\figurenum{9}
\includegraphics[angle=-90,scale=0.65]{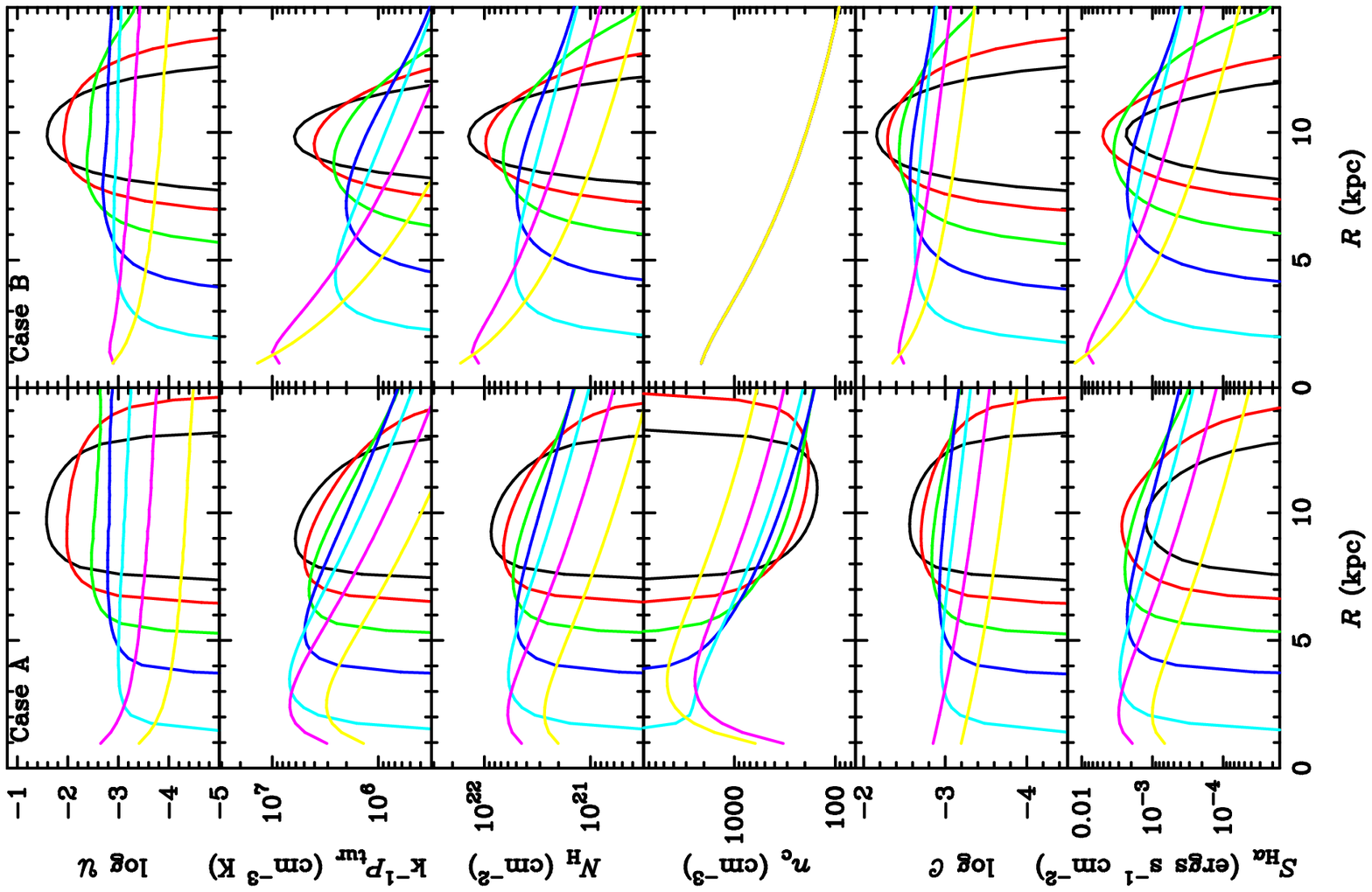}
\figcaption{\footnotesize The photoionization models for given gaseous disks. From the top to bottom
panels are ionization parameter $\calu$ (at the illuminated face), $k^{-1}P_{\rm tur}$, column density
$N_{\rm H}$ ($=0.5\siggas/m_p$ for photoionzation model), number density of clouds $n_c$, the filling
factor $\calc$ and the H$\alpha$ brightness surface $S_{\rm H\alpha}$ of the gaseous disk. Lines with
different colors correspond to the times in Figure 7 and 8.}
\label{fig9}
}

It should be noted that the present ionization parameter $\calu$ does not depend on star numbers
($N_*$) of the cluster if since $\varsigma$ or $f$ are independent of $N_*$. The hot gas pressure
reads $P_{\rm hg}=n_{\rm hg}kT_{\rm hg}$, where $k$ is the Boltzmann constant, $n_{\rm hg}$ and
$T_{\rm hg}$ are the density and temperature of the hot gas, respectively. Stellar wind supplies
the hot gas through heating of shocks produced by the interaction between
the winds and medium of the inner surface. The power of the stellar winds is given by
$L_{\rm w}=E_{\rm w}/t_{\rm w}$, where $E_{\rm w}\sim N_*M_{\rm w}V_{\rm w}^2$, $M_{\rm w}$ is the
mass of stellar wind, $V_{\rm w}$ is the wind speed and $t_{\rm w}$ is the timescale of the wind
replenishment. Considering the balance of the shock heating and free-free cooling, $t_{\rm w}$
should be equal to the free-free cooling timescale
($t_{\rm ff}$). This yields $t_{\rm w}=t_{\rm ff}\propto n_{\rm hg}^{-1}T^{1/2}$, and
$L_{\rm w}\propto N_*M_{\rm w}V_{\rm w}^2n_{\rm hg}T_{\rm hg}^{-1/2}$, which is balanced by the
free-free cooling $\propto n_{\rm hg}^2T_{\rm hg}$. We then have the hot gas density
$n_{\rm hg}\propto N_*M_{\rm w}V_{\rm w}^2T_{\rm hg}^{-3/2}$ and its thermal pressure
$P_{\rm hg}\propto n_{\rm hg}T_{\rm hg}\propto N_*M_{\rm w}V_{\rm w}^2T_{\rm hg}^{-1/2}
\propto N_*L_*^{1.7}V_{\rm w}^2T_{\rm hg}^{-1/2}$, where we use $M_{\rm w}\propto L_*^{1.7}$
(Pellegrini et al. 2007) and $L_*$ is star luminosity. The radiation pressure reads
$P_{\rm rad}\propto N_*L_*$ from $N_*$ stars. It is expected that the parameter
$\varsigma=P_{\rm hg}/P_{\rm rad}\propto L_*^{0.7}$ is independent of $N_*$, but
dependent of $L_*$ inside the clump. From the simple argument, $\calu$ is independent of $N_*$,
but depends on the IMF and the density of the clumps.

\subsection{Emission from the gaseous disks}
We have 50 histories of star formation rates for these 50 belts. This can be justified by the
condition
\begin{equation}
t_{\rm adv}\ge \min(t_{\rm MS},t_{\rm SF}),
\end{equation}
where $t_{\rm MS}$ is the lifetime of the hydrogen main sequence stars, and $t_{\rm SF}$ is the
timescale of star formation. We find that 50 belts are accurate enough for the current calculations.
For the $j-$belt at time $t$, the ionizing luminosities are calculated from the star formation rates
as
\begin{equation}
\Delta \dot{M}_*(R_j,t)=2\pi R_j \Delta R\sigsfr(R_j,t),
\end{equation}
where $R_j$ is the radius of the $j-$belt with a width $\Delta R$. We get the total spectroscopic luminosity
of the each belt from BC03, i.e., $\Delta \dot{M}_*(R_j,t)\rightarrow\Delta L_{\nu}^*(R_j,t)$. Given
the ionizing luminosity, ionization parameter and density of the ionized gas, CLOUDY produces the line
luminosity $\Delta L_l(R_j,t)$. The surface brightness of emission lines is given by
\begin{equation}
S_l(R_j,t)=\frac{\Delta L_l(R_j,t)}{2\pi R_j\Delta R}.
\end{equation}
This equation produces the relation between ratio of emission lines and $S_{\rm H\alpha}$, which can
be compared with observations to test the theoretical model. We list all the necessary equations for
the calculations of photoionization of the gaseous disk.

{
\centering
\figurenum{10}
\includegraphics[angle=-90,scale=0.6]{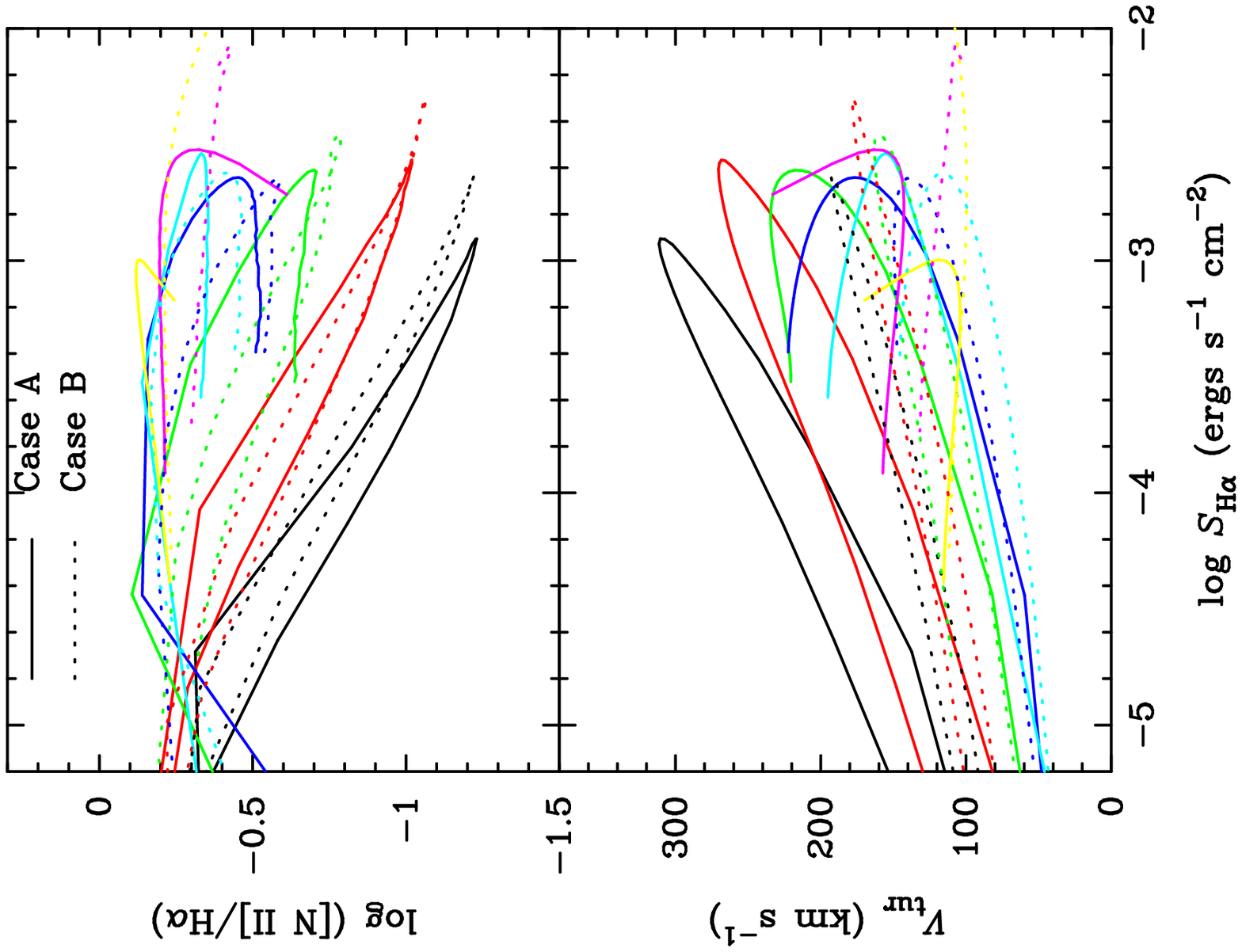}
\figcaption{\footnotesize The upper panel shows the relation of the line ratio of $\NII/{\rm H}\alpha$
with the H$\alpha$ brightness. The lower panel shows the relation of the turbulence velocity and the
brightness. Loops with different colors are obtained along the radius of the gaseous disk at the times
corresponding to that in Figure 7 and 8.}
\label{fig10}
}

\figurenum{11}
\begin{figure*}[t]
\centerline{\includegraphics[angle=-90,scale=0.7]{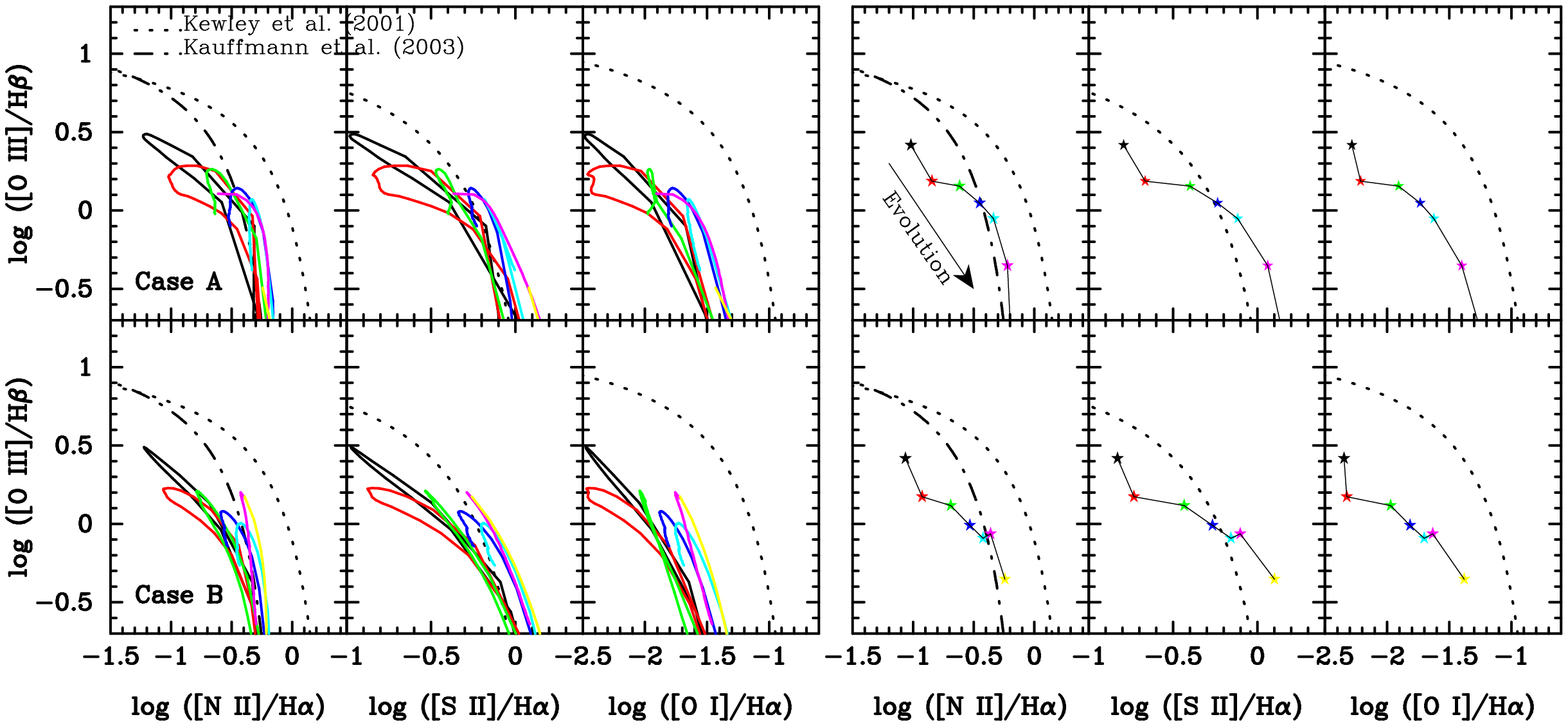}}
\figcaption{\footnotesize The BPT diagram for the dynamical structure given by Figure 2-3 and 7-9.
{\em Left:} Lines with different colors are at the
same time in Figure 7 and 8. Loops with different colors are obtained along the radius of the gaseous
disk at the times. {\em Right:} stars with different colors correspond
to the line ratios integrating the whole galaxies show the evolutions with time. It is clear that the galaxies
are evolving along the arrow marked in the BPT diagram.}
\end{figure*}
\label{fig11}

\subsection{Numerical results}
Figure 9-11 show the photoionization models corresponding to the dynamical structures for Case A and B.
The ionizing fluxes follow the star formation rates of the gaseous disk. Figure 9 shows the parameters
of photoionization model along the radius. The ionization parameter $\calu$ displays a synchrotron
evolution with disk's dynamics, but holds a constant over the gaseous disk. The ionizing photons are
shown in Figure 7 and 8. We find that the gaseous disks are quite clumpy at a level of $\calc\sim 10^{-3}$.
H$\alpha$ brightness is calculated from equation (40) according to the photoionization model. The
H$\alpha$ ring is evolving following the evolution of the dynamical structure and star formation rates.

Figure 10 shows the relation between line ratio $\NII$/H$\alpha$ and H$\alpha$ brightness in the upper panel.
This plot actually reflects the relation of the dynamical structure and photoionization. Clearly the trend
is consistent with the observations as shown in Figure 11 in Lehenert et al. (2009). Emission line brightness
is proportional to density and ionization parameter, namely, $S_l\propto n_e\calu$ while the line ratios
depends on the $\calu^{-1/2}$. It is thus expected that there is an anti-relation between
$S_l$ and line ratios for given the density. Star formation rates roughly determine the ionization parameter
as well as the gas pressure inside the clouds (equation 36). Temperature is thus high as a consequence of the
photoionization shown by the $k^{-1}P-$panel in Figure 9.  The model produces a trend of turbulent velocity
with H$\alpha$ brightness as shown in the lower panel of Figure 10, which is consistent with Figure 6 in
Lehnert et al. (2009). Generally, the brighter H$\alpha$ ring the more intensive star formation, and hence
stronger turbulence.

\figurenum{12}
\begin{figure*}
\centerline{\includegraphics[angle=-90,scale=0.7]{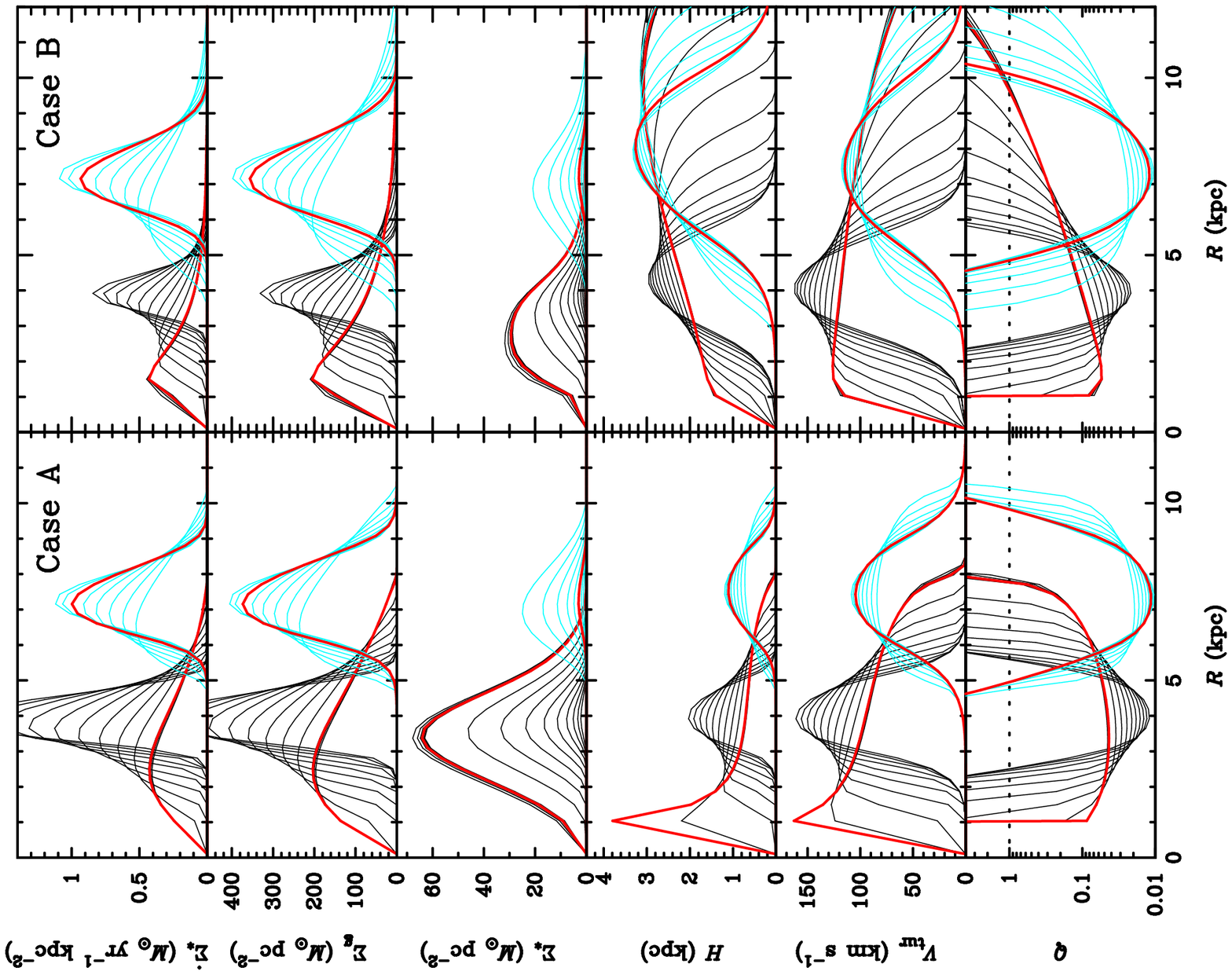}}
\figcaption{\footnotesize
Evolution of for BX482 calculated from both Case A (right) and B (left) models.
The parameters used are listed in Table 3. The thin black and blue lines represent evolution of the first and
the second merged rings, respectively. The red lines are the current status of the rings at the observational
times. The second ring evolves 3Myr since the minor merger. The interval of the two successive rings is
$\sim 10^8$yr. The profile of turbulent velocities agrees with the observed rings (Figure 3 in Genzel et al.
2008).}
\end{figure*}
\label{fig12}

We calculate emission line ratios for the BPT diagram (Baldwin, Phillips \& Terlevich 1981)\footnote{There
are extensive calculations of the BPT diagram by adjusting the ionization parameter $\calu$ (e.g. Dopita
et al. 2000, Kewley et al. 2001), but
there are no calculations combined with the dynamical structure of star forming galaxies.
According to the results from Dopita et al. (2000) and Kewley et al. (2001), the line ratio relations are
mainly determined by the ionization parameter $\calu$.} to test if there
is evolutionary trend of star forming galaxies.  In principle, the line ratios depend on the ionization
parameter and metallicity mainly. Evolution and gradient of metallicity are complicated,
but we fix the metallicity for simplicity to show how the line ratios evolve with ongoing star formation.
We find that the ratio of $\OIII/{\rm H}\beta$ is decreasing
with $\NII/{\rm H}\alpha$, $\SII/{\rm H}\alpha$ and $\OI/{\rm H}\alpha$ in term of increases of $\calu$.
Figure 11 shows the BPT diagram of the gaseous disks for a series of given times. Since $\calu$ is a function
of radii as shown by Figure 9, the line ratios show open loops in the BPT diagram. Considering the open loops
are significantly overlapped, we integrate the whole galaxies to get the global ratios for the evolutionary
trend of the galaxies. Obviously, the galaxies are evolving as shown in {\em right} part of Figure 11.

It is worth examining the evolutionary trend presented here, but it needs a large sample which allows us
to estimate  $\calu$ through star formation rates. Starting from SDSS data, we
group the sample in light of $\calu$ and replot the BPT diagram to test the evolutionary trend. The results
will be
carried out in a separate paper. We would like to point out that we fix the metallicity for all radius
and duration of evolution of galaxies. The real evolution of galaxies should be more complicated than
that of pure star formation. It would be very motivated if we incorporate metallicity evolution with
dynamics.

\subsection{Uncertainties of the present model}
The main uncertainty of the photoionization is the parameter $\varsigma$ (or $f$), which appears
in the ionization parameter $\calu$ in equation (37). The ionization structure should be actually
self-consistently determined by the dynamics controlled by the stars and stellar evolution inside the clumps,
however, this is extremely complicated and beyond the scope of the present paper. We invoke the observational
constraints from M17 cluster (Pellegrini et al. 2007) to simplify the complex problem. We show
$\varsigma \propto L_*^{0.7}$, implying that $\calu$ depends on IMF through both
$\varsigma$ and $\langle \nu\rangle$. Though the uncertainties
of this point, the main characters of the photionized structure of the gaseous disk remain. Future work on
the structure and photoionization is worth exploring.
Finally, we do not include the role of the mechanism energy of SNexp in line emission through shock
heating as argued by Kewley et al. (2001). The role of shock heating in line emission mainly depends
on the efficiency $(\xi)$. We will consider the role of SNexp heating for line emission in the future
paper.

\section{Applications: BX 482 and BX 389}
A few dozen of high$-z$ galaxies are well observed currently in the SINS sample. They are often less
evolved. As applications of the present model, we choose two galaxies from
SINS sample observed through the integral-field spectroscopy. BX 482 and BX 389 are
selected for applications, whose dynamical structure and emission lines are presented by Genzel et al.
(2008) and Lehnert et al. (2009), respectively. BX 482 is rotating star-forming outer ring/disk dominated
galaxy while BX 389 is central bulge/inner disk dominated (Genzel et al. 2008). On the other hand, they are
located in the star forming
area in the BPT diagram (Figure 9 and 10 in Lehnert et al. 2009). This sets up additional constraints on
the details of both dynamical structure and emission of the two galaxies.

Generally, turbulence velocity, H$\alpha$ surface brightness and $\SII\lambda6716/\SII\lambda6731$ can be
measured from observations. For the free parameters listed in Table 1, we first estimate the orders of each
parameters used in the model. Gas surface density is mainly determined by the H$\alpha$ surface brightness,
we have the estimation of $\siggas$. The parameter $\xi$ can be estimated from equation (15) or (19) in
term of $V_{\rm tur}$ for Case A and B, respectively. Given the estimation of $\xi$ and $n_c$,
$\alpha$ can be evaluated by equation (33). The timescale of dynamical evolution due to SNexp-driven viscosity
is then given by $t_{\rm adv}=R^2/\nu$. Though some parameters degenerate, we try to estimate them separately.
The averaged mass density is roughly $n_c\sim 1200{\rm cm^{-3}}$ estimated from the line ratio of
$\SII\lambda6716/\SII\lambda6731$ values (Lehnert et al. 2009). Given the dynamical structure, we calculate
the photoionization model.

\subsection{BX 482}
BX 482 is a ring-like star forming galaxy with small bulge ($\sim 2\times 10^{10}\sunm$) shown by H$\alpha$
image, which has a ring radius of $R_{\rm ring}=7.0\pm 0.8$kpc with a width $\Delta R\sim 1$ kpc and a
turbulence velocity $V_{\rm tur}=100$km~s$^{-1}$ (Genzel et
al. 2008). The galaxy is undergoing intensive star formation with a rate\footnote{The uncertainties of the
star formation rates  is larger than 50\% in SINS sample (see Table 2 in Genzel et al. 2008). Here we take
the uncertainties of 50\% for both BX 389 and BX 482.} of $\sim 140\pm 70\sunmyr$. It shows
two Gaussian profiles for the bulge and ring. The rotation curve measured by H$\alpha$ line
show a similar shape of rotation curve with equation (11). We set $R_c=3.5\rm{kpc}$ and $V_c=300\kms$ for the
rotation curve in this object. The star formation rate in the small bulge is of
$\dotsfr_{\rm bulge}\sim 35\pm 17\sunm~{\rm yr}^{-1}$
(potentially with a disk geometry) while the star formation rate is of
$\dotsfr_{\rm disk}\sim 105\pm 51\sunm~{\rm yr}^{-1}$ in the
ring (estimated from Figure 3 in Genzel et al. 2008). This indicates that star formation is still ongoing in
both the bulge and the disk of the galaxy yet. Clearly, it is necessary to have two successive minor mergers
to explain the evolution of BX 482. The initial parameters of the two rings are given in Table 3.

\vglue 0.42cm
{
\centering
\figurenum{13}
\includegraphics[angle=-90,scale=0.65]{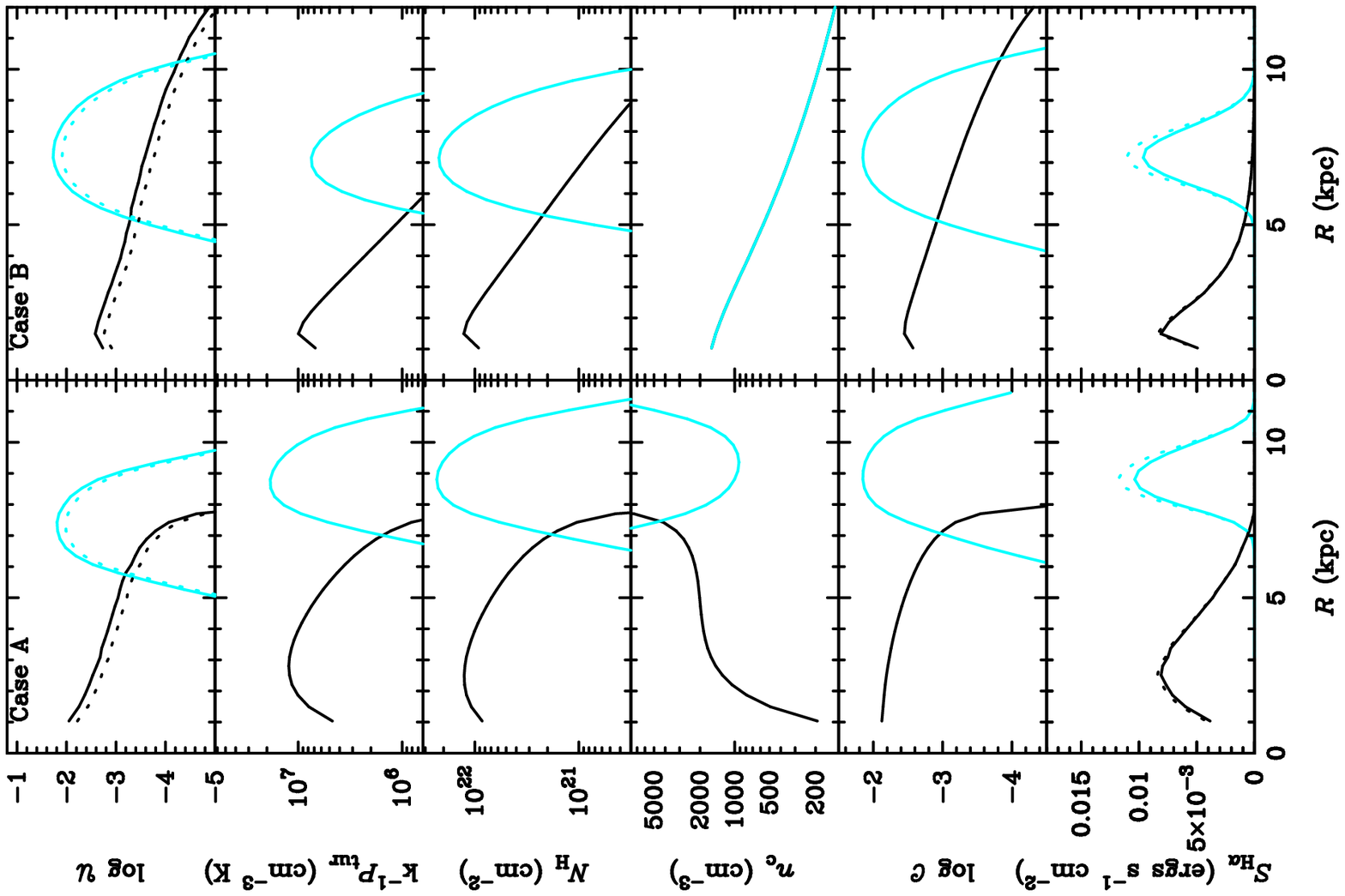}
\figcaption{\footnotesize Photoionization model for BX 482 for the given structure in Figure 12. The black
and blue lines represent the current status of the evolving rings. The dotted and solid lines are for
$f=0.4,0.6$, respectively. We point out that the density $n_c\propto \Sigma_{\rm gas}^{-0.8}$ as shown by
eq. (33) for Case A whereas the two lines overlap together for Case B.
}
\label{fig13}
}

Figure 12 shows the dynamical structure of the models of Case A and B for BX 482. Generally, the models can
explain the observed dynamical structures. $\xi$ for the second ring
is smaller than that for the first in term of the smaller turbulence velocity. We integrate $\sigsfr$ over
the two rings and have the star formation rates. For Case A, we have $\dotsfr_{\rm bulge}\sim 33\sunmyr$
in the bulge and $\dotsfr_{\rm disk}\sim 92\sunmyr$ in the galactic disk. This agrees with the observed star
formation rates. For Case B, we have $\dotsfr_{\rm bulge}\sim 17\sunmyr$ and $\dotsfr_{\rm disk}\sim 90\sunmyr$,
which are still consistent with the observed within the uncertainties. The stellar surface $\Sigma_*$, composed
of the most old stars, is obtained from the integration of the star formation rates.
The height of the gaseous disk is of 2kpc in the late phase of
the evolution, which is rough in agreement with the scaled height observed ($\sim 1.6$kpc). The profile
of turbulence velocity of Case A is similar to the observed (compared with Figure 3 in Genzel et al. 2008)
and better than that of Case B (which is too flat). Clearly the clouds of the gaseous disk are self-gravity
dominated in light of $Q<1$ driving star formation. The first ring has been immigrated into the bulge,
which is still growing. The fate of the second ring is uncertain. It may stay at the current radius form
a stellar disk for the given $\xi$ in Table 3. If $\xi$ is time-dependent and increases with time (discussed
in \S5), the ring is estimated that the second ring will immigrate into the bugle in about $\sim 10^8$ yrs
as shown by the first and the third panel of Figure 12.

Figure 13 shows photoionization models and emission from the star forming disks at the evolutionary epoch
of $10^8$yrs. The two gaseous rings have very different $\calu$ since they are at different evolutionary
epoch as well as other physical parameters. We find the gaseous disk is very clumpy
as shown by $\calc$, which is of $10^{-2}\sim 10^{-3}$. Equation (40) can produce the H$\alpha$ emission
which allows us to compare with H$\alpha$ image. The last panel shows the surface brightness of H$\alpha$
emission. We find that the results are in agreement with observed as shown in the distribution of the
surface brightness (see Figure 3 in Genzel et al. 2008). We also plot the line ratios in the BPT diagram
in Figure 16, which is consistent with the observations.

\figurenum{14}
\begin{figure*}
\centerline{\includegraphics[angle=-90,scale=0.7]{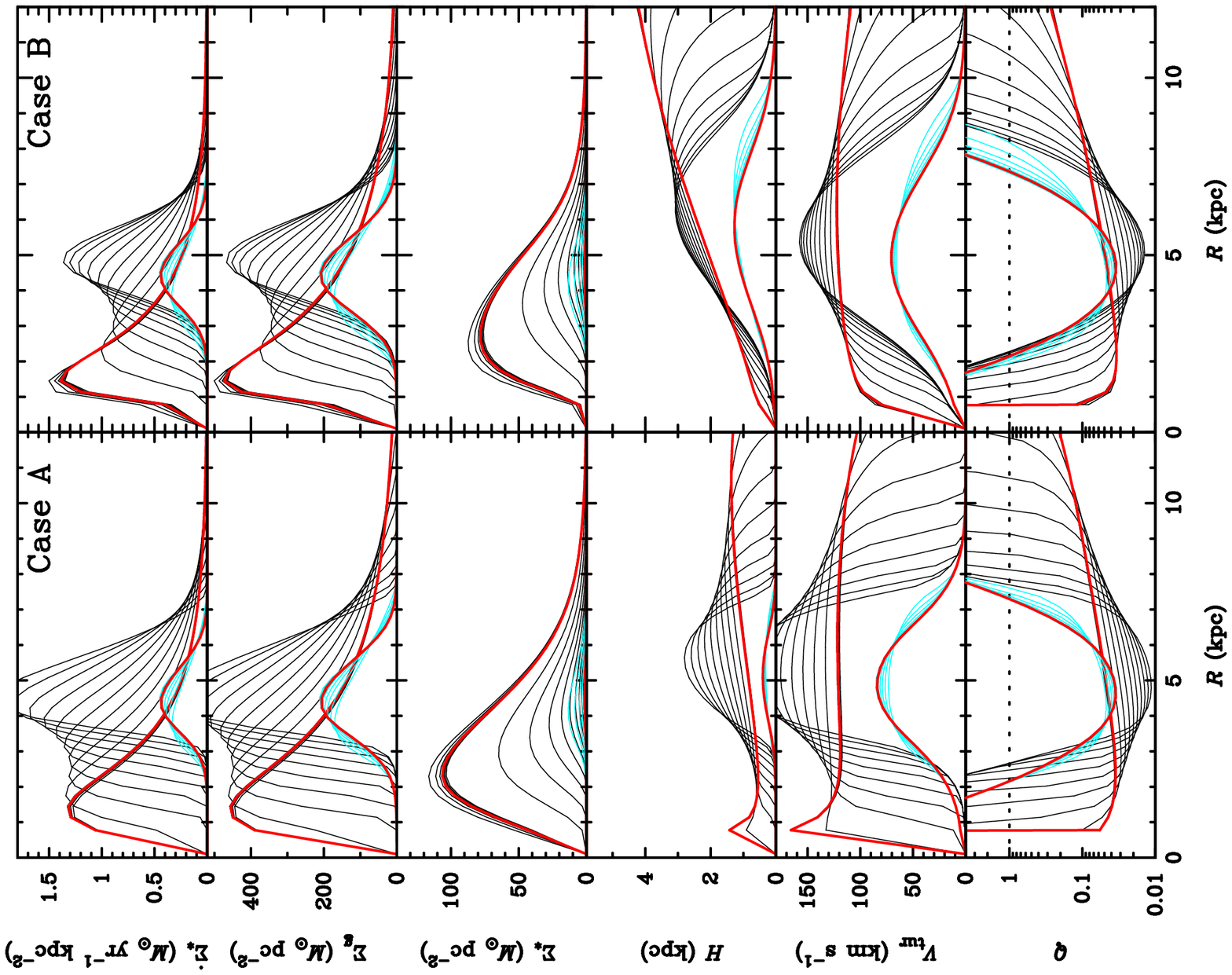}}
\figcaption{\footnotesize Dynamical structure and evolution of BX 389, which show bulge-dominated structure.
The lines have similar meanings but for different initial conditions as in Figure 12. The interval of the
two rings are $10^8$yr. The turbulent velocity profile can be compared with Figure 7 in Genzel et al. (2008).
}
\end{figure*}
\vglue 0.3cm
\label{fig14}

\begin{table}%[h!]
{\footnotesize\begin{center}
\caption{Values of parameters in applications}
\begin{tabular}{llllll}\hline\hline
&\multicolumn{2}{c}{BX 482}& &\multicolumn{2}{c}{BX 389}\\
\cline{2-3}\cline{5-6}
Parameter &Case A&Case B& &Case A&Case B\\ \hline
$V_c$        &300       &300       & &315       &315 \\
$R_c$        &3.5       &3.5       & &1.8       &1.8 \\
$f$          &0.6, 0.4  &0.6, 0.4  & &0.6, 0.4  &0.6, 0.4 \\
$\alpha$     &0.3       &0.2       & &0.4       &0.2 \\ \hline
\multicolumn{6}{c}{Initial parameters of the first gaseous ring}\\ \hline
$\xi_1/10^{-2}$    &1.3    &1.2  & &2.4     &1.0 \\
$R_{0,1}$          &4.0    &4.0    & &5.0       &5.0 \\
$\Delta_{R,1}$     &0.5    &0.5    & &0.8       &1.0 \\
$\Sigma_{0,1}$     &230    &120    & &320       &250 \\
$M_{0,1}$          &2.5    &1.3    & &5.6       &4.5 \\
$M_{0,1}/M_{\rm dyn}$&0.2  &0.1    & &0.4       &0.3 \\ \hline
\multicolumn{6}{c}{Initial parameters of the second gaseous ring}\\ \hline
$\xi_2/10^{-2}$    &0.8    &0.38 & &1.5      &0.3 \\
$R_{0,2}$          &7.3    &7.3    & &4.5       &4.5 \\
$\Delta_{R,2}$     &0.8    &0.8    & &1.0       &1.0 \\
$\Sigma_{0,2}$     &120    &120    & &120       &120 \\
$M_{0,2}$          &4.3    &4.3    & &1.8       &1.8 \\
$M_{0,2}/M_{\rm dyn}$&0.3   &0.3    & &0.1       &0.1 \\ \hline
\end{tabular}\end{center}}
\end{table}

\subsection{BX 389}
As another illustration, we apply the present model to BX 389, which is a bulge-dominated star forming
galaxy with a weak ring or disk at a scale of $R_{\rm ring}=4.4\pm 0.5$kpc and a
turbulence velocity $V_{\rm tur}=180$km~s$^{-1}$. This object is very different from BX 482. The total
star formation rates are $150\pm 75\sunmyr$, of which $\sim 100\pm 50\sunmyr$ in the bulge and
$\sim 50\pm 25\sunmyr$ in the ring. We set $R_c=1.8\rm{kpc}$ and $V_c=315\rm{km/s}$ to describe the
BX 482's rotation curve (see Fig. 1). Other parameters are listed in Table 3. Two rings are assumed in
modeling the dynamics and photoionization images. Since the star formation rates in the bulge dominate
over in the weak ring or the disk (Genzel et al. 2008) and turbulence velocity in BX 389 is faster than
that in BX 482, this implies that the dynamical evolution of the gas ring should be
quite fast, namely, the parameter $\xi$ should be relatively higher than that in BX 482.
This can be justified by equation (15) or (19). We set the initial mass of the first ring
is $M_{0,1}=5.6\times 10^{10}\sunm$ at radius $R_{0,1}=5.0$kpc, which is mainly determined by the present
star formation rate in the dynamical time. The black lines describe the dynamic evolution of the first
ring. We find that the ring fast evolved inside the region of the bulge with a timescale of $10^8$yr
since $\xi$ is higher than that in BX 482. The second ring is injected after 0.1Gyr of evolution of the
first ring.
The initial mass is determined by the surface density of H$\alpha$, but the dynamical age is unknown.
The main difference from the first ring is the initial mass, and other parameters are quite similar.
Figure 14 shows the dynamical structure and evolution of the two rings. The star formation rates in the
bulge and disk are estimated as $\dotsfr_{\rm bulge}\sim 75\sunmyr$ and $\dotsfr_{\rm disk}\sim 27\sunmyr$
for Case A, respectively, and $\dotsfr_{\rm bulge}\sim 70\sunmyr$ and $\dotsfr_{\rm disk}\sim 27\sunmyr$
for Case B, respectively. The ratio of star formation rates is roughly one third in weak disk to the
bulge, consistent with the observed. We note that the first merger ratio is about 1:2.5, which is slightly
larger than the definition of a minor merger. This could be called a small major merger, which still validate
the present model approximately.

Photoionization model is shown by Figure 15, which corresponds to the dynamical structure in Figure
14 for both Case A and B. The line ratios from the photoionization model are plotted in Figure 16 as
the BPT diagram. The current model can produce the main features of dynamical structure of the two
rings. The uncertainties of the model result in the dynamical age of the thick disk or weak ring,
but the future track of the ring is predicted by the theoretical model.
The spectroscopy for the weak ring is not separated from the total observationally.

In summary, we show that emission and dynamical structure are simultaneously reproduced by the
present model. The SNexp, as a key ingredient in the model, is driving the evolution
of galaxies.
It should be noted that the merged masses in both BX 389 and BX 482 are at a level of $10^{10}\sunm$.
Such large mass accretions ($>10^{10} M_\odot$) must be accreted over roughly $\sim 10^8$ years, which
is consistent with the gas accretion rates onto galaxies at these masses. If they are accreted on time
scales shorter than this, they must be due to mergers, not cold gas accretion. The two galaxies favor
mergers.

{
\centering
\figurenum{15}
\includegraphics[angle=-90,scale=0.65]{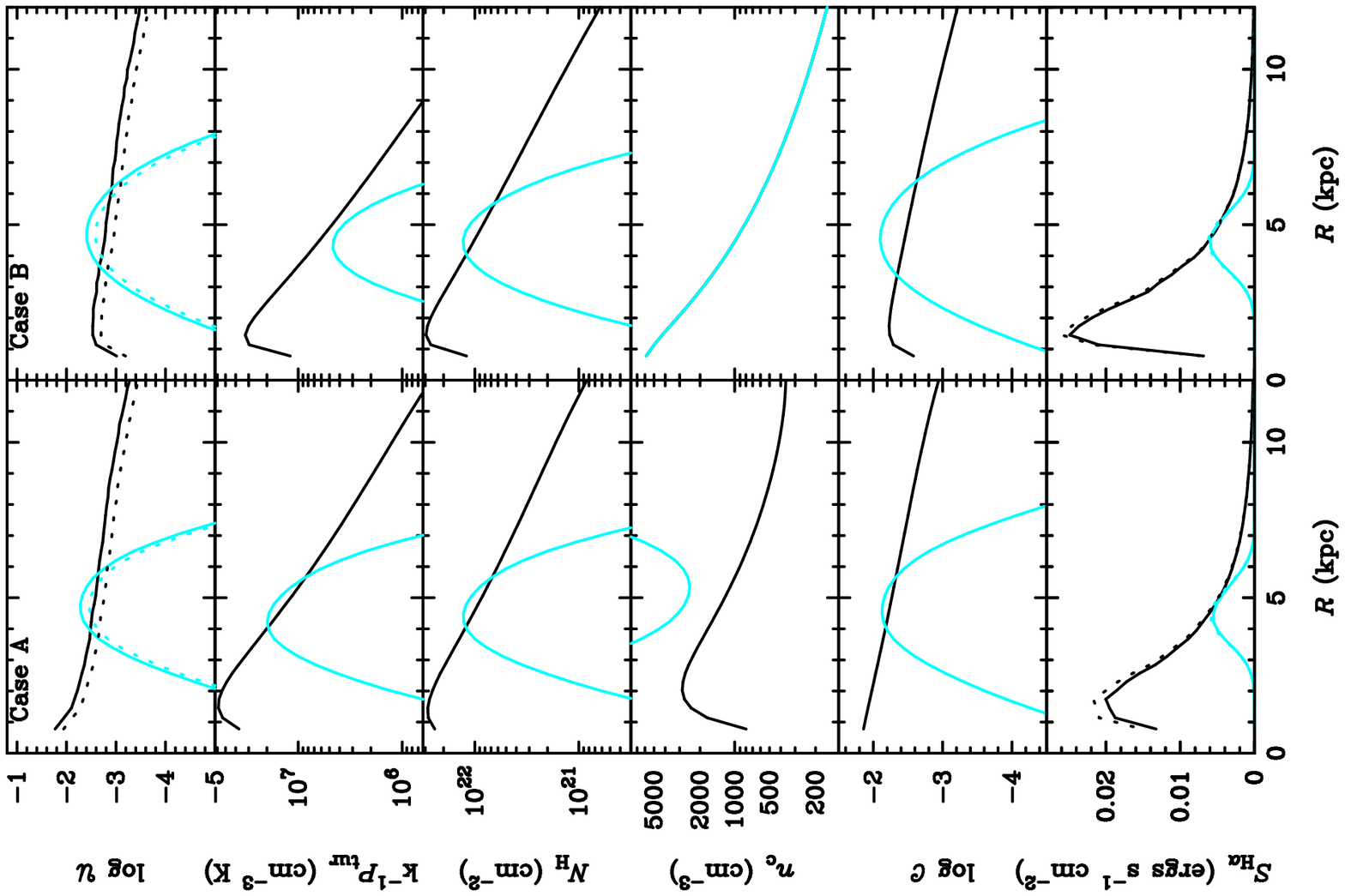}
\figcaption{\footnotesize Photoionization model for BX 389 for the given structure in Figure 13. The black
and blue lines represent the current status of the evolving rings. The dotted and solid lines are for
$f=0.4,0.6$, respectively. Similar properties of $n_c$ can be found in the caption of Fig. 13.
}
\label{fig15}
\vglue 0.3cm
}

\figurenum{16}
\begin{figure*}
\centerline{\includegraphics[angle=-90,scale=0.6]{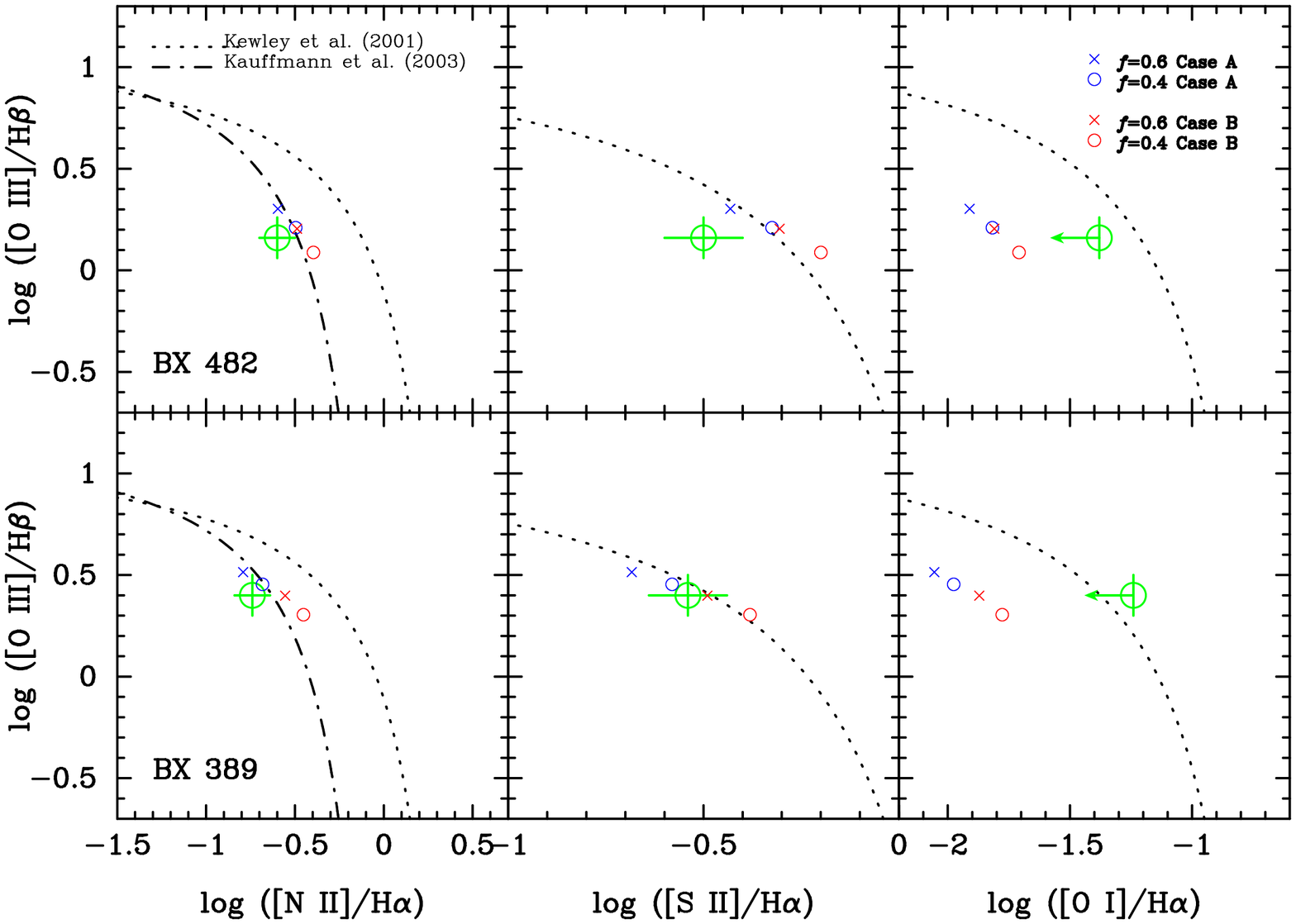}}
\figcaption{\footnotesize BPT diagram, the green circles are observation datas for the two galaxies. We
take the line ratios (green symbols) from Figure 9 and 10 in Lehnert et al. (2009), which were estimated
in a square aperture of $0.8''\times 0.8''$ centered on the peak of the H$\alpha$ emission. The model
points (blue and red symbols) are integrated in the same regions.}
\label{fig16}
\vglue 0.3cm
\end{figure*}

\section{Discussions}
We have presented detailed calculations of dynamical structure and emission of the gaseous disk.
Both rough estimates and numerical results of gaseous disk structures show that
the parameter $\xi$ has strong influence on the structure and evolution. In this section, we briefly
discuss several potential factors influencing $\xi$, such as, the roles of galactic winds, stellar
evolution and initial mass function.  The delayed SMBH activities and cosmological evolution as
further implications of the present model are also discussed briefly.

\subsection{Influence of galactic winds}% Further implications}
Galactic-scale winds generally exist (e.g. Chen et al. 2010) and are
thought to play an important role in evolution of galaxies.
The present star formation rate density is much higher than the critical threshold of
0.1$\sunmyrkpcc$, above which is powerful enough to drive vigorous galactic-scale outflows in local
galaxies (Heckman 2003). The outflows are expected to play an important role in rendering a starburst
"self-regulation" (Dopita \& Ryder 1994; Silk 1997) supported by evidence of blue wings in rest-frame
optical emission lines (e.g. Nesvadba et al. 2007). Its influence should be incorporated into equations
(1-3), even in the energy equation (6). Furthermore, broad emission lines of $z\sim 2$ galaxies have
been found, which show H$\alpha$ FWHM of 1500$\kms$ (Shapiro et al. 2009). Such a broad component is
thought to origin from galactic winds. The influence of stellar winds on the gaseous disk results in
the gas loss and hence affects on the star formation. Transportation of angular momentum of the gas
by SNexp-driven turbulence is stressed in this paper, which is supported by triggering of supermassive
black holes hosted at centers of galaxies (Chen et al. 2009, see also Hobbs et al. 2010). Obviously,
galactic winds will significantly lead to losses of mass and angular momentum, and hence change the
evolutionary track of the gaseous disks. Likely, the peak in $\siggas-$track appears in a further outer
region in the presence of strong winds, even disappears.

The well-known metallicity-mass relation (e.g. Panter et al. 2008) could be a result of galactic winds
since less massive galaxies have shallower potential so to allow a stronger outflows. It is expected
that outflows have influence on the photoionization through metallicity.

\subsection{IMF, stellar evolution and metallicity}
There is increasing evidence for the cosmological evolution of IMF (e.g. Kroupa 1999; Dave 2008), implying
that the $\xi$ may undergo cosmic evolution. Top-heavy IMF appears in high redshift and may lead to more
violent evolution of high$-z$ galaxies, and AGN activities. Cosmic evolution of galaxies should include
the role of IMF evolution in the dynamic equations. Additionally, the correlation between
metallicity and galaxy mass indicates that more massive galaxies are metal-richer (Tremonti et al. 2004;
Brooks et al. 2007; Finlator \& Dave 2008), namely, SNexp rates may undergo at different levels in galaxy
evolution in light of star formation efficiency depending on galaxy mass.  Obviously, massive galaxies
already underwent many episodes of star formation in the past, resulting in metal-rich.
The present paper focuses on the evolution of structure on a dynamical timescale of galaxies,
though the IMF determines the metallicity in a complicated way (beyond the scope of the present paper),
we do not include the cosmic evolution of IMF and fix it for all radii in the present calculations.
%Different metallicity at different radii and time might change the location in the BPT diagram.
IMF determines the SNexp rates, which control the dynamical structure of galaxies, leading to the
metallicity distribution in galaxies. As a result of photoionization, metallicity govern the location
of the galaxies in the BPT diagram.  Future work by
including the evolution of metallicity will provide more strong constraints on theoretical model and
unveil the histories of mergers.

We simply assume $\xi$ as a constant with radius and time within a dynamical timescale of galaxies,
but it should be a function of radius
due to the stellar mass function and stellar evolution. For massive stars,
the hydrogen burning lifetime is given by $t_{\rm MS}=4.5\left(M_*/40\sunm\right)^{-0.43}$Myr. For
early phase of galaxy formation, the minimum mass of stars should be greater than
$M_c\ge 40 \Delta t_{\rm 5Myr}^{-2.3}\sunm$, we have $\xi\propto \Delta t_{\rm 5Myr}^{3.1}$ after
inserting $M_c$ into equation (28), where $\Delta t_{\rm 5Myr}=\Delta t/5{\rm Myr}$. These massive
stars will dramatically change the very early phase of the gaseous ring. SNexp rates as a function
of time change $\xi$ by a factor of a few, may significantly change the results presented here.
This is the reason why we use a smaller one $\xi$ for the second ring which is only of a few $10^6$yrs
in both BX 482 and BX 389 (see Table 3).

\subsection{Inside-out growth of galaxies}
It is well-known that inside-out growth of galaxy disks drives the gradients of averaged age of stars,
colors and metallicity (Bouwens et al. 1997), however, M33 has a truncated disk at $R\sim 8$kpc (Ferguson
et al. 2007) and the outer disk show an age increase with radius (Barker et al. 2007a, 2007b; Williams
et al. 2009), indicating that the real growth of galaxy disks could be more complicated than the canonic
version of inside-out growth. The present model only deals with the single episode of a minor merger or 
cold flows, and predicts three different fates of the merged gas as shown in Figure 2{\em a}-2{\em c} 
and Figure 3{\em a}-3{\em c}. For a cosmological evolution of galaxies, their growth histories contain 
various populations of minor mergers or cold flows, and form composite gradients of age and metallicity. 
As interesting applications of the present model, histories of galaxy growths will be formulated from 
the population synthesis of minor mergers or cold flows.

We use a presumed rotation curve in the paper, namely, we separate the gaseous disk from the stellar
components of host galaxies. However, the inside-out growth of galaxies is governing the potential
on cosmic timescale. We do not include the potentials of the gas itself and the newly formed
stars. This is valid for a single episodic activity of star formation. For a cosmic evolution of entire
galaxies, we should change the rotation curves with inclusion of the newly formed stars in the last
episodic activity and then get the history of star formation in the entire galaxies step by step.
We would like to point out the rotation curves (equation 11) can be replaced by other forms
under different potentials of galaxies. The present numerical scheme can conveniently include this.
Future applications to the large SINS sample should modify the rotation curves for more complicated
structures and emission.

\subsection{Coevolution of SMBHs and galaxies}
Delivering gas into the centers of galaxies by SNexp will lead to coevolution of supermassive black
holes and galaxies, which is established for local galaxies (Magorrian et al. 1998; Richstone et al.
1998) and may undergo cosmic evolution in high-$z$ galaxies and quasars (e.g. Coppin et al. 2008).
Star forming galaxies at high redshift must undergo violent evolution: gas assembly, star formation,
feedback from star formation and mergers and black hole activity. As a consequence, galaxies are
growing fast at this epoch and may suffer from AGN feedback, showing increase of AGN duty cycle with
redshifts (Wang et al. 2006; 2008). Supermassive black holes are triggered by mergers (e.g. Roos 1985;
Carlberg 1990; Wyithe \& Loeb 2002), especially minor mergers (Li et al. 2009) with a natural delay
at a level of the transportation timescale following a star burst (Chen et al. 2009; Paper I; Netzer
2009; Kawakatu \& Wada 2008; Davies et al. 2009; Wild et al. 2010). A unified model for galactic
evolution and AGNs is required to describe the coevolution, in which feedback from star formation
and AGN are included. The $\eta-$equation statistically describes the coevolution of SMBHs and
galaxies (Wang et al. 2009a), but it should incorporate dynamics of galaxies and accretion through
transportation of angular momentum driven by SNexp turbulence of young massive stars. In such a
scenario, a delayed inflow to the central SMBHs is triggered in this way (Di Matteo et al. 2005;
Hopkins et al. 2006; Chen et al. 2009; Netzer 2009; Kumar \& Johnson 2010). We can in principle
establish the dynamics of coevolution.

\section{Conclusions}
We have a detailed extension of Paper I to investigate the roles of SNexp incorporated into
the dynamics, for structures and emission of evolving gaseous disk with given potential of host galaxies.
Dynamical properties
of such a gaseous disk is totally different from the classical gaseous disk with a constant viscosity,
which simply diffuses into space. We show that SNexp can efficiently transport gas into
more inner regions and form a stellar ring or thick stellar disk. SBHM activity is then triggered
with a lag of $\sim 10^8\left({\rm SFR}/10^2\sunmyr\right)^{-0.9}$yr relative to starbursts.
A stellar ring or disk forms, or immigrates into bulges depending on the efficiency $\xi$. For the case
with high viscosity, the stellar ring formed from the secular evolution of the gaseous disk will immigrate
into the bulge with dynamical timescale due to stellar friction. Bulges and SMBHs are jointly growing
for high viscosity case whereas only galactic disks are growing for low viscosity. The model is able
to produce the observed structures of high$-z$ galaxies, including height, velocity dispersion and
young stellar components.

Emission from the photoionized gaseous disk has been calculated through the simple stellar synthesis.
UV photons are efficiently photoionizing the gas surrounding the star clusters and showing emission
lines, such as $\OIII$, $\NII$, H$\alpha$ and H$\beta$. The model predicts relations of line ratios
and H$\alpha$ brightness, BPT diagram. We apply the models to SINS sample, yielding the main features
of the high$-z$ galaxies. The consistency of the theoretical
models with the observational properties sheds light of the roles of SNexp in galaxy evolution.

Future work to improve the present model are extensively discussed, especially the stellar evolution,
galactic winds, initial mass function and the geometry of ionized clumps. We stress the factors of
influencing the efficiency $\xi$. When it is a function of time
and radius, the structure and evolution of the gaseous disk will be changed somehow.
The present model for single episodic event can be extended for successive multiple minor mergers
or cold flows, whose initial conditions are different from each other. The composite
models by including all episodes could explain the metallicity and age distributions in galaxies
in the frame of the inside-out growth, or more complicated observational results.

\acknowledgements The authors are grateful to the referee for a very constructive and helpful
report significantly improving the manuscript. The paper is dedicated to JMW's father, who
unfortunately passed away in June during the completion of the revised version of the paper,
and expresses the infinite grief. We appreciate the stimulating discussions
among the members of the IHEP AGN group. The research is supported by NSFC-10733010 and 10821061,
CAS-KJCX2-YW-T03, and 973 project (2009CB824800).

\newpage

{\appendix

\section{Numerical Scheme}
The non-linear partial differential equations (18) and (22) can be solved numerically by
predictor-corrector methods (Lu \& Guan 2003). For the standard form of the nonlinear differential
equation.
\begin{eqnarray}
\left\{
\begin{array}{lll}
\frac{\D\partial^2Y}{\D\partial x^2}=k_1(x,Y)\frac{\D\partial{Y}}{\D\partial{t}}+k_2(x,Y),&~~~ &(0<x<1,~{\rm and~} t>0),\\ \\
Y(x,0)=g(x),&~~~ & (0\leqslant x \leqslant 1),\\ \\
Y(0,t)=\mu_1(t), \frac{\D\partial Y(1,t)}{\D\partial x}=\mu_2(t),&~~~ &(t \geqslant0),
\end{array}
\right.
\end{eqnarray}
where $\mu_1(t)$ and $\mu_2(t)$ are functions of boundary conditions. We divide the disk into $J$
($j=0,1,...,J$) rings with a width $h=dx$. For simplicity, we define two operators as
\begin{equation}
\delta Y_j=Y_{j+\frac{1}{2}}-Y_{j-\frac{1}{2}}, ~~~~~\mu Y_j=\frac{1}{2}\left(Y_{j+\frac{1}{2}}+Y_{j-\frac{1}{2}}\right).
\end{equation}
For the $j-$th ring at time $n\tau$, we have the predictor
\begin{equation}
\frac{1}{h^2}\delta^2Y_j^{n+\frac{1}{2}}=k_1\left(x_j, Y_j^n\right)\frac{Y_j^{n+\frac{1}{2}}-Y_j^n}{\tau/2}
                                           +k_2\left(x_j, Y_j^n\right),~~~~~(j=1,..., J-1),
\end{equation}
and the corrector
\begin{equation}
\frac{1}{2h^2}\delta^2(Y_j^n+Y_j^{n+1})=
k_1\left(x_j,Y_j^{n+\frac{1}{2}}\right)\frac{Y_j^{n+1}-Y_j^n}{\tau}+k_2\left(x_j,Y_j^{n+\frac{1}{2}}\right),
~~~~~(j=1,...,J-1),
\end{equation}
where $\tau=dt$ is the interval of the time. The initial condition is given by
\begin{eqnarray}
Y_j^0=g(jh), &~~~~~(j=1,2,...,J-1).
\end{eqnarray}
The inner boundary condition is given by
\begin{equation}
Y_0^n=\mu_1(n\tau).
\end{equation}
The outer boundary can be expressed in two ways for the predictor and corrector
\begin{equation}
\frac{1}{h}\mu\delta Y_J^n=\mu_2\left(\frac{n\tau}{2}\right), ~~~~~{\rm and}~~
\frac{1}{2h}\mu\delta\left(Y_J^n+Y_J^{n+1}\right)=\mu_2\left(\frac{n\tau}{2}\right),
\end{equation}
respectively. Setting $j=J$ in equations (A3) and (A4) at the outer boundary, we have
\begin{equation}
\frac{1}{h^2}\delta^2Y_J^{n+\frac{1}{2}}=k_1\left(x_J, Y_J^n\right)\frac{Y_J^{n+\frac{1}{2}}
                                           -Y_J^n}{\tau/2}+k_2\left(x_J, Y_J^n\right),
\end{equation}
\begin{equation}
\frac{1}{2h^2}\delta^2\left(Y_J^n+Y_J^{n+1}\right)=
k_1\left(x_J,Y_J^{n+\frac{1}{2}}\right)\frac{Y_J^{n+1}-Y_J^n}{\tau}+k_2\left(x_J,Y_J^{n+\frac{1}{2}}\right).
\end{equation}
Combining with (A7), (A8) and (A9), we can obtain $Y_J$ and $Y_{J-1}$ by eliminating $Y_{J+1}$. We get
$Y_j^{n+\frac{1}{2}}$ from (A3) and put it in (A4) for $Y_j^{n+1}$, and so on through the alternative
direction implicit (ADI) method.
}

\end{document}